\documentclass{article}

\usepackage[numbers]{natbib}         
\usepackage[colorlinks]{hyperref}    
\usepackage[english]{babel} 
\usepackage{amssymb}
\usepackage{amsmath}
\usepackage{txfonts}
\usepackage{mathdots}
\usepackage[classicReIm]{kpfonts}
\usepackage[dvips]{graphicx} 
\usepackage[a4paper, portrait, margin=1in]{geometry}
\usepackage{tabularx}
\usepackage{longtable}
\usepackage{multirow}
\usepackage{booktabs}
\usepackage{lscape}
\usepackage[labelsep=period]{caption}
\usepackage{makecell}
\begin{document}
	\setlength{\parindent}{0pt}
	\setlength{\parskip}{1ex}
	
	\textbf{\Large Deep Learning in MRI-guided Radiation Therapy: A Systematic Review}
	
	\bigbreak

	Zach Eidex$^{1,2}$, Yifu Ding$^{1}$, Jing Wang$^{1}$, Elham Abouei$^{1}$, Richard L.J. Qiu$^{1}$, Tian Liu$^{3}$,
	Tonghe Wang$^{4}$ and Xiaofeng Yang$^{1,2}$*

	$^{1}$Department of Radiation Oncology and Winship Cancer Institute, Emory University, Atlanta, GA
	
	$^{2}$School of Mechanical Engineering, Georgia Institute of Technology, Atlanta, GA
	
	$^{3}$Department of Radiation Oncology, Icahn School of Medicine at Mount Sinai, New York, NY
	
	$^{4}$Department of Medical Physics, Memorial Sloan Kettering Cancer Center, New York, NY

	\bigbreak
	\bigbreak
	\bigbreak

	\textbf{*Corresponding author: }
	
	Xiaofeng Yang, PhD
	
	Department of Radiation Oncology
	
	Emory University School of Medicine
	
	1365 Clifton Road NE
	
	Atlanta, GA 30322
	
	E-mail: xiaofeng.yang@emory.edu

	\bigbreak
	\bigbreak
	\bigbreak
	\bigbreak
	\bigbreak
	\bigbreak

	\textbf{Abstract}

	\textbf MRI-guided radiation therapy (MRgRT) offers a precise and adaptive approach to treatment planning. Deep learning applications which augment the capabilities of MRgRT are systematically reviewed. MRI-guided radiation therapy offers a precise, adaptive approach to treatment planning. Deep learning applications which augment the capabilities of MRgRT are systematically reviewed with emphasis placed on underlying methods. Studies are further categorized into the areas of segmentation, synthesis, radiomics, and real time MRI. Finally, clinical implications, current challenges, and future directions are discussed.

	\bigbreak
	\bigbreak
	
	\textbf{keywords:} MRI-guided, Radiation Therapy, Radiotherapy, Deep Learning, Review.

	\noindent 
	\section{ INTRODUCTION}
	
	Recent innovations in magnetic resonance imaging (MRI) and deep learning are complementary and hold great promise for improving patient outcomes. With the advent of the Magnetic Resonance Imaging Guided Linear Accelerator (MRI-LINAC) and MR-guided radiation therapy (MRgRT), MRI allows for accurate and real-time delineation of tumors and organs at risk (OARs) that may not be visible with traditional CT based plans.\cite{RN1} Deep learning methods augment the capabilities of MRI by reducing acquisition times, generating electron density information crucial to treatment planning, and increasing spatial resolution, contrast, and image quality. In addition, MRI auto-segmentation and dose calculation methods greatly reduce the required human effort on tedious treatment planning tasks, enabling physicians to further optimize treatment outcomes. Finally, deep learning methods offer a powerful tool in predicting the risk of tumor recurrence and adverse effects. These advancements in MRI and deep learning usher in the era of fully adaptive radiation therapy (ART) and the MRI-only workflow.\cite{RN2}
	
	Deep learning methods represent a broad class of neural networks which derive abstract context through millions of sequential connections. While applicable to any imaging modality, these algorithms are especially well suited to MRI due to its high information density.\cite{RN3} Deep learning demonstrates state of the art performance over traditional hand-crafted and machine learning methods but are computationally intensive and require large datasets. For MRI and other imaging tasks, convolutional neural networks (CNNs), built on local context, have traditionally dominated the field. However, advancements in network architecture, availability of more powerful computers, large high-quality datasets, and increased academic interest have led to rapid innovation. Especially exciting are the rapid adaptation of cutting edge recurrent, attention, and self-attention methods which continue to improve upon and even replace CNNs.
	
	Deep learning techniques can be organized according to their applications in MRgRT in the following groups: segmentation, synthesis, radiomics (classification), and real-time/4D MRI. Segmentation methods automatically delineate tumors, organs at risk (OARs), and other structures. However, deep learning approaches face challenges when adapting to small tumors, multiple organs, low contrast, and differing ground truth contour quality and style. These challenges differ greatly depending on the region of the body, so segmentation methods are primarily organized by anatomical region.\cite{RN4} 
	
	Synthesis methods are best understood by their input and output modalities. Going from MRI to CT, synthetic CT (sCT) provides accurate attenuation information not apparent in MRI, augmenting the information of  co-registered CT images. In an MRI-only workflow, sCT avoids registration errors and the radiation exposure associated with traditional CT.\cite{RN5} In addition, synthetic relative proton stopping power (sRPSP) maps can be generated to directly obtain dosimetric information for proton radiation therapy.\cite{RN6} The dosimetric uncertainty can be further enhanced with deep learning dose calculation methods which greatly reduce inference time and could yield lower dosimetric uncertainties compared to traditional Monte Carlo (MC) methods. Synthetic MRI (sMRI) generated from CT, is appealing by combining the speed and dosimetric information of CT with MRI’s high soft tissue contrast. However, CT’s lower soft tissue contrast makes this application much more challenging, but sMRI has still found success in improving CT-based segmentation accuracy.\cite{RN7, RN9, RN8} Alternatively, there are rich intra-modal applications by generating one MRI sequence from another. For example, the spatial resolution of clinical MRI can be increased by predicting a higher resolution image\cite{RN11, RN10} and applying contrast can be avoided with synthetic contrast MRI.\cite{RN12} 
	
	Radiomics represents an eclectic body of works but can be divided into studies which classify structures in an MRI image\cite{RN13} or prognostic models which use MR images to predict treatment outcomes such as tumor recurrence or adverse effects.\cite{RN15, RN14} Deep learning methods in real-time and 4D MRI overcome MRI’s long acquisition time and the low field strengths of the MRI-LINAC by reconstructing images from undersampled k-space\cite{RN16}, synthesizing additional MRI slices\cite{RN17}, and exploiting periodic motion to improve image quality\cite{RN18}. 
	
	In this review, we systematically examine studies that apply deep learning to MRgRT, categorizing them based on their application and highlighting interesting or important contributions. We also discuss future trends in deep learning and MRgRT.

	\noindent 
	\section{LITERATURE SEARCH}
	
	This systematic review surveys literature which implements deep learning methods and MRI for radiation therapy research. “Deep learning” is defined to be any method which includes a neural network directly or indirectly. These include machine learning models and other hybrid architectures which take deep learning derived features as input. Studies including MRI as at least part of the dataset are included.  Studies must list their purpose as being for radiation therapy and include patients with tumors. Studies on immunotherapy and chemotherapy without radiation therapy are excluded. Conference abstracts and proceedings are excluded due to an absence of strict peer review. 
	
	The literature search was performed on Pubmed on December 31, 2022, with the following search criteria in the title or abstract: “deep learning and (MRI or MR) and radiation therapy”. This search yielded 335 results. Of these results, 197 were included based on manual screening using the aforementioned criteria. 78 were classified as segmentation, 81 as synthesis, 24 as radiomics (classification), and 14 as real-time or 4D MRI. There is inevitably some overlap in these categories. In particular, studies which use sMRI for the purposes of segmentation are classified as synthesis and papers which deal with real-time or 4D MRI are placed in Section 6: Real-Time and 4D MRI. Figure 1 shows the papers sorted by category and year. Compared to other review papers, this review paper is more comprehensive in its literature search and is the first specifically on the topic of deep learning in MRgRT. In addition, this work uniquely focuses on the underlying deep learning methods as opposed to their results. Figure 2 shows technical trends in deep learning methods implementing 3D convolution, attention, recurrent, and GAN techniques.

	\begin{figure}
		\centering
		\noindent \includegraphics*[width=6.50in, height=4.20in, keepaspectratio=true]{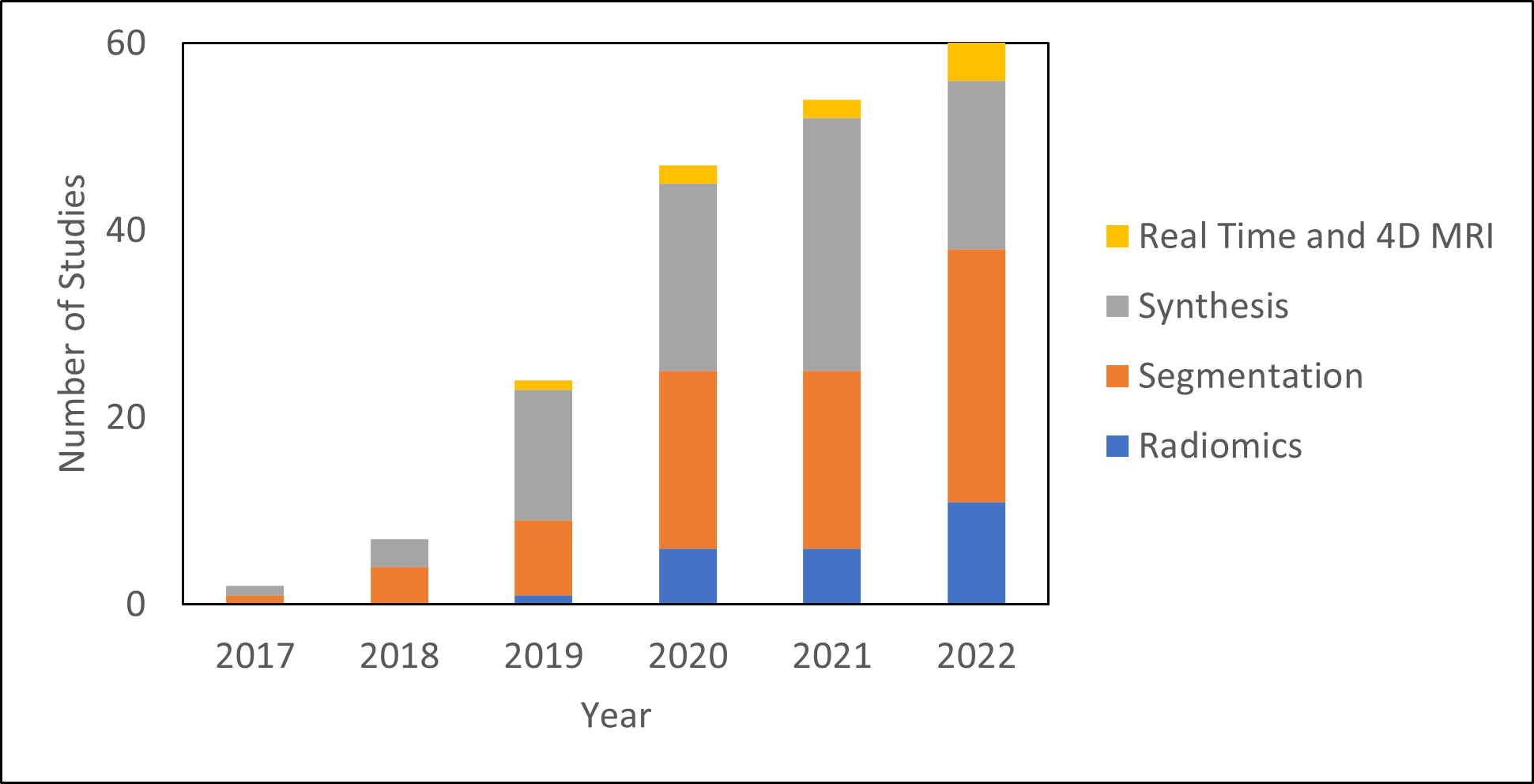}
		
		\noindent Figure 1. Number of deep learning studies with applications towards MRgRT per year by category.
	\end{figure}

	\begin{figure}
		\centering
		\noindent \includegraphics*[width=6.50in, height=4.20in, keepaspectratio=true]{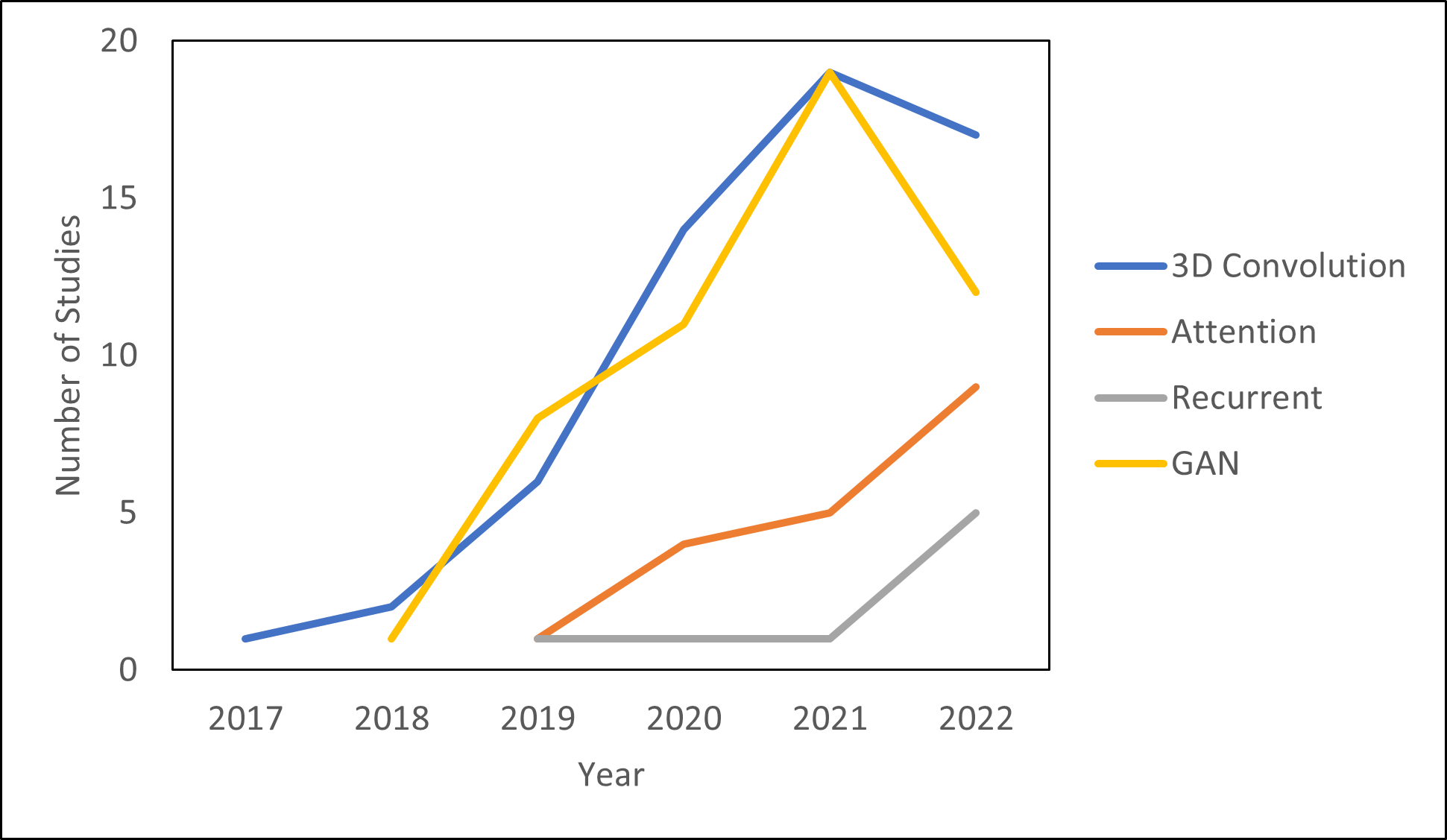}
		
		\noindent Figure 2. Technical Trends in Deep Learning.
	\end{figure}

	\noindent 
	\section{IMAGE SEGMENTATION}
	
	Contouring (segmentation) in MRgRT is the task of delineating targets of interest on MR images which can be broadly divided into distinct categories: contouring of organs at risk and other anatomical structures expected to receive radiation dose and contouring of individual tumors. Contouring is typically performed by dosimetrists, physicists, and physicians. Both tumor and multi-organ segmentation suffer from intra- and inter- observer variability.\cite{RN19} MRI does not capture the true extent of the tumor volume, as well as poorly defined boundaries and similar structures like calcifications lead to institutional and intra-observer variability. Physician contouring conventions and styles further complicate the segmentation task and lead to inter-observer variability.\cite{RN21, RN20} Multi-organ segmentation is mostly challenged by the large number of axial slices and OARs which make the task tedious and prone to error. Automated solutions to MRI segmentation have been proposed to reduce physician-workload and provide expert-like performance.
	
	Since the application of CNNs to MRI-based segmentation in 2017\cite{RN22}, fully convolutional networks (FCNs) have outperformed competing atlas-based and hand-crafted auto-segmentation methods, often matching the intra-observer variability among physicians\cite{RN23}. FCNs employ convolutional layers which are trained to detect patterns in either nearby voxels or feature maps output from previous convolutional layers. In contrast with traditional CNNs, FCNs forgo densely connected layers. This design choice enables voxel-wise segmentation, allows for variable sized images, and reduces model complexity and training time. Different types of convolutions include atrous and separable convolutions. Atrous convolutions sample more sparsely to gain a wider field of view and can be mix-and-matched to capture large and small features in the same layer. Separable convolutions divide a 2D convolution into two 1D convolutions to use fewer parameters for similar results. By connecting multiple convolutional layers together with non-linear activation functions, larger and more abstract regions of the input image are analyzed to form the encoder. For pixelwise segmentation, the final feature map is expanded to the original image resolution through a corresponding series of transposed convolutional layers forming the decoder. All FCNs include pooling layers to conserve computational resources whereby the resolution of feature maps is reduced by choosing the largest (max-pooling) or average local pixel.\cite{RN24}
	
	To evaluate performance, various evaluation metrics are employed with the Dice similarity coefficient (DSC) being the most prevalent. The DSC is defined in equation 1 (Eq 1) as the overlap between the ground truth physician contours and the predicted algorithmic volumes with a value of 0 corresponding to no overlap and 1 corresponding to complete overlap. Mathematically, it is defined as follows where VOLGT is the ground truth volume and VOLPT is the predicted volume\cite{RN25}:

	\begin{equation} 
		DSC=\frac{2|VOLGT \land VOLPT|}{|VOLGT|+|VOLPT|}   
	\end{equation}
	
	Additional metrics include the Hausdorff distance\cite{RN25} which measures the farthest distance between two points of the ground truth and algorithmic volumes, volume difference\cite{RN26}, which is simply the difference in volumes, and the Jaccard Index\cite{RN27}, which is similar to the DSC and measures the overlap between VOLPT and VOLGT relative to their combined volumes. A discussion of these metrics is found in Müller et al.\cite{RN28} However, performance between datasets must be evaluated with caution due to high inter-observer variation between physicians and dataset quality.
	
	The properties of MRI datasets have driven innovation. Multiple MRI sequences, with and without contrast, are often available. To capture all data, the different sequences are co-registered and input as multiple channels yielding multiple segmentations. These segmentations are combined to produce a final segmentation using an average, weighted average, or more advanced method. To account for MRI’s high through-plane resolution relative to its in-plane resolution, 3D convolutional layers are often utilized to capture features not apparent in 2D convolution. However, 3D convolutions are computationally expensive, so numerous 2.5D architectures have been proposed.\cite{RN30, RN31, RN29} In a 2.5D architecture, adjacent MRI slices are input as channels, and 2D convolutions are performed. It is also common to see new papers forgo the 3D convolution to save resources for new computationally intense methods. An unfortunate fact is that high-quality MRI datasets are often small. To remedy this, data augmentation methods such as rotating and flipping the MR images are ubiquitous. In addition, the generation of synthetic images to increase dataset size and generalizability is an exciting field of research.\cite{RN275} Public datasets and competitions have also helped in this regard. For example, the Brain Tumor Segmentation Challenge (BraTS) dataset\cite{RN32}, updated since 2012, has been a primary contributor to brain segmentation progress, spawning the popular DeepMedic framework\cite{RN33}. Another approach for small datasets is transfer learning. In transfer learning, a model is trained on a large dataset, and then retrained on a smaller dataset with the idea that many of the previously found features are transferable.\cite{RN34}
	
	Advances from the field of natural language processing (NLP) have had a tremendous impact on segmentation tasks. Recurrent neural networks (RNNs) are defined by the output of their node being connected to the input of their node. To avoid an infinite loop, the output is only allowed to connect to its input a set number of times. This property allows for increased context and the ability to handle sequential data which is especially important in language translation. Applied to CNNs, each recurrent convolutional layer (convolution + activation function) is preformed multiple times which creates a wider field of view and more context with each subsequent convolution. However, recurrent layers can suffer from a vanishing gradient problem. Long short-term memory blocks (LSTM) solve this by adding a “forget” gate which forgets irrelevant information. In addition, LSTMs are more capable of making long range connections.  Similar to the LSTM gate, the gated recurrent unit (GRU) has an update and reset gate which decide which information to pass on and which to forget. Both LSTM and GRU also have bidirectional versions which pass information forward and backwards.\cite{RN35, RN36} Relative performance between the LSTM and GRU gates are situational with the GRU gate being less computationally expensive.\cite{RN37}
	
	A major issue faced in MRI-segmentation can be characterized as “the small tumor problem”. Small structures like tumors or brachytherapy fiducial markers represent a small fraction of the total MRI volume, where CNNs can struggle to find them or be confused by noise. Further exacerbating the problem is that applying a deep CNN to whole MR images consumes extensive computational resources, so the MRI must be downsampled. In this case, the downsamplying is very likely to cause small tumors to be missed entirely. One of the simplest ways to improve performance is to alter the loss function. Standard loss functions are cross-entropy and dice loss which seek to maximize voxel wise classification accuracy and overlap between the predicted and ground truth contours, respectively. These can be modified to achieve higher sensitivity to small structures at the expense of accuracy. Focal loss is the cross-entropy loss modified for increased sensitivity\cite{RN38} and Tversky loss does the same for the dice loss.\cite{RN39} In addition, borders of the contours are the most important part of the segmentation, so boundary loss functions seek to improve model performance by placing increased emphasis on regions near the contour edge.\cite{RN72, RN42}  Another approach to solve the problem, albeit at the expense of long-range context, is with two stage networks. In the first stage, regions of interest (ROIs) are identified, and target structures are then contoured in the ROIs in the second stage. Notable efforts include Mask R-CNN\cite{RN40} and Retina U-Net\cite{RN41} which implement convolution-based ROI sub-networks with advanced correction algorithms. Seqseg instead replaces the correction algorithms with a reinforcement learning based model.\cite{RN42} An agent is guided by a reward function to iteratively improve the conformity of the bounding box. Seqseg reported comparable performance with higher bounding box recall and intersection over union (IoU) compared to Mask R-CNN.
	
	Related developments from NLP are the concepts of attention and the transformer. In terms of MRI, attention is the idea that certain regions of the MRI volume are more important to the segmentation task and should have more resources allocated to them. ROI schemes can then be defined as a form of hard attention by only considering the region around a tumor. A version of soft attention would weight the region around the tumor heavily and process the information in high resolution but also give a smaller weighting to nearby organs and process it in lower resolution.\cite{RN43} In practice, attention modules include a fully connected feedforward neural network to generate weights between a feature map of the encoder and a shallower feature map in the decoder. These weights are improved upon through backpropagation of the entire network to give higher representational power to contextually significant areas of the image. This fully connected network can also be replaced with other models such as the RNN, GRU, or LSTM.\cite{RN44} If the same feature map is compared with itself, this is called self-attention and is the basis for the transformer architecture.\cite{RN45} The transformer can be thought of as a global generalization of the convolution and can even replace convolutional layers. The advantages of the transformer are explicit long-range context and the transformer’s multi-head attention block allows for attention to be focused on different structures in parallel. However, transformers require more data to train and can be very computationally expensive. Such computational complexity can be remedied by including convolutional layers in hybrid CNN-transformer architectures\cite{RN46}, by making long range connections between voxels sparse,\cite{RN47} or by implementing more efficient self-attention models like FlashAttention.\cite{RN277}
	
	From the field of neuroscience, deep spiking neural networks (DSNNs) attempt to more closely model biological neurons by connecting neurons with asynchronous time dependent spikes instead of the continuous connections between neurons of traditional neural networks. Potential advantages include lower power use, real-time unsupervised learning, and new learning methods. However, these advantages are only fully realized with special neuromorphic hardware, are difficult to train, and currently lag conventional approaches. For these reasons, they are currently only represented by one paper in this review.\cite{RN48}
	
	Many new models for MRI segmentation have been created by modifying U-Net. U-Net derives its name from its shape which features convolutional layers in the encoder and transposed convolutional layers in the decoder. Its main innovation, however, is its long-range skip connections between the encoder and decoder.  Dense U-Net densely connects convolutional layers in blocks\cite{RN49}, ResU-Net includes residual connections\cite{RN50}, Retina U-Net is a two-stage network, RU-Net includes recurrent connections, R2U-Net adds residual recurrent connections\cite{RN51}. Attention modules have also been added at the skip connections.\cite{RN53, RN52} The aforementioned networks were all designed with 2D convolutions but can be modified to include 3D convolutions. Both V-Net\cite{RN54} and nnUNet\cite{RN55} were designed with 3D convolutional layers with nnUNet additionally automating preprocessing and learning parameter optimization. Pix2pix uses U-Net as the generator with a convolutional discriminator (PatchGAN)\cite{RN56}. Other state-of-the-art architectures include Mask R-CNN, DeepMedic, and DeepLabV3+.\cite{RN57} Mask R-CNN is a two-stage network with a ResNet backbone. Mask Scoring RCNN (MS-RCNN) improves upon Mask R-CNN by adding a module which penalizes ROIs with high classification accuracy but low segmentation performance\cite{RN58}. DeepMedic, designed for brain tumor segmentation, is an encoder-only CNN which inputs a ROI and features two independent row-resolution and normal resolution channels. These channels are joined in a fully connected convolutional layer to predict the final segmentation. The convolutions in the encoder-only style reduce the final segmentation map dimensions compared to the original ROI (25x25x25 vs 9x9x9 voxels). DeepLabV3+ leverages residual connections and multiple separable atrous convolutions. Xception improves upon the separable convolution by reversing the order of the convolutions and including ReLU blocks after each operation for non-linearity.\cite{RN59}

	\noindent 
	\subsection{Brain}
	
	Largely unaffected by patient motion and comprised of detailed soft tissue structures, the brain is an ideal site to benchmark segmentation performance for MRI and represents the dominant category in MRI segmentation research. Unique to brain MRI preprocessing is skull stripping, where the skull and other non-brain tissue are removed from the image. This can significantly improve results, especially for networks with limited training data.\cite{RN60} Shown in table 1, the majority of the studies focus on segmenting different brain tumors such as glioma, Glioblastoma Multiforme (GBM), and metastases. A small minority of studies focuses on OARs like the hippocampus. Advancements in brain segmentation have come, in large part, from the yearly Multimodal Brain Tumor Image Segmentation Benchmark (BraTS) challenge, which includes high quality T1-weighted (T1W), T2-weighted (T2W), T1-contrast (T1C), and T2 -Fluid-Attenuated Inversion Recovery (FLAIR) sequences with the purpose of segmenting the whole tumor (WT), tumor core (TC), and enhancing tumor (ET) volumes. The WT is defined as the entire spread of the tumor visible on MRI; The ET is the inner core which shows significant contrast compared to healthy brain tissue, and the TC is the entire core including low contrast tissue. The most popular architectures are DeepMedic, created for the BraTS challenge, and U-Net. 
	
	Notable efforts in the BraTS challenge include Momin et al achieving an exceptional WT dice score of .97 ±.03 with a Retina U-Net based model and mutual enhancement strategy. In their model, Retina U-Net finds a ROI and segments the tumor. This feature map is fed into the classification localization map (CLM module) which further classifies the tumor into subregions. The CLM shares the encoding path with a segmentation module, so classification and segmentation share information and are improved iteratively.\cite{RN61} Huang et al focuses on correctly segmenting small tumors. Based on DeepMedic, the method incorporates a prior scan and custom loss function, the volume-level sensitivity–specificity (VSS), which rates and significantly improves the metastasis sensitivity and specificity to segment small brain metastases.\cite{RN62} Another paper improves small tumor detection by 2.5 times compared to the standard dice loss by assigning a higher weight to small tumors.\cite{RN63} Lee et al takes the novel approach of using standard dice loss for the first 40 epochs and changes to Tversky loss for the final 20 epochs to specify sensitivity and specificity. \cite{RN64} Both Tian et al\cite{RN65} and Ghaffari et al\cite{RN66} utilize transfer learning datasets to cope with limited data. Pan et al includes a two-stage U-Net model with residual and attention blocks. \cite{RN67} Ahmadi et al achieves competitive results in the BraTS challenge with a DSNN. \cite{RN48} 
	
	{\small\tabcolsep=3pt  
	\begin{longtable}{ >{\bfseries\footnotesize}llllllll }
		\caption{Brain Segmentation Studies}
		\label{tab:my-table}\\
		\hline
		Study & Year & Target & \begin{tabular}[c]{@{}l@{}}Network\\    \\ Architecture\end{tabular} & \begin{tabular}[c]{@{}l@{}}Network\\    \\ Features\end{tabular} & Imaging Modalities & Patient Number & DSC \\ \hline
		Simon et al\cite{RN39} & 2022 & AVM & U-Net &  & TOF MRA, T1WC, T2W & 23 & \begin{tabular}[c]{@{}l@{}}arteries: 0.86\\    \\ veins: 0.91\\    \\ brain: 0.98\\    \\ CSF: 0.91\end{tabular} \\ \hline
		Tian et al\cite{RN63} & 2022 & GBM & U-Net & 3D & T1C & 20 & 0.94 ± 0.012 \\ \hline
		Bouget et al\cite{RN66} & 2022 & GBM & AGU-Net & Attention & T1W, FLAIR & 2134 & 0.86 ± 0.17 \\ \hline
		Momin et al\cite{RN59} & 2022 & Glioma & Retina U-Net & \begin{tabular}[c]{@{}l@{}}BRATS\\    \\ ROI\\    \\ 3D\end{tabular} & T1, T1C, T2W, FLAIR & 369 & \begin{tabular}[c]{@{}l@{}}WT: 0.97 ± 0.03\\    \\ TC: 0.90 ± 0.13\\    \\ ET: 0.77 ± 0.22\end{tabular} \\ \hline
		Ma et al\cite{RN67} & 2022 & Meningioma & CNN & \begin{tabular}[c]{@{}l@{}}Residual\\    \\  Recurrent \\    \\ Attention\end{tabular} & T1C & 551 & 0.89 \\ \hline
		Mi et al\cite{RN69} & 2022 & Temporis & U-Net &  & 3.0T T1C & 132 & 0.89 \\ \hline
		Huang et al\cite{RN60} & 2022 & Tumor & DeepMedic & 3D & T1C & 176 & 0.81 \\ \hline
		\begin{tabular}[c]{@{}l@{}}Chartrand\\    \\ et al\cite{RN61}\end{tabular} & 2022 & Tumor & U-Net & 3D & T1W & 530 & \begin{tabular}[c]{@{}l@{}}2.5-6mm:0.68\\    \\ > 10mm:0.86\end{tabular} \\ \hline
		Yoo et al\cite{RN23} & 2022 & Tumor & U-Net & 2.5D & T1C & 65 & 0.75 \\ \hline
		Ghaffari et al\cite{RN64} & 2022 & Tumor & Dense U-Net & \begin{tabular}[c]{@{}l@{}}Dense\\    \\ 3D\end{tabular} & T1W, T1C, T2W, T2-FLAIR & 15 & \begin{tabular}[c]{@{}l@{}}WT: 0.83\\    \\ TC: 0.77\\    \\ ET: 0.60\end{tabular} \\ \hline
		Bouget et al\cite{RN70} & 2021 & GBM & nnUNet & \begin{tabular}[c]{@{}l@{}}BRATS\\    \\ 3D\end{tabular} & T1C, T1W & 1887 & 0.87 ± 0.15 \\ \hline
		Pan et al\cite{RN65} & 2021 & HC & U-Net & \begin{tabular}[c]{@{}l@{}}Attention\\    \\ Residual\\    \\ ROI\\    \\ 3D\end{tabular} & 1.5 T T1W & 235 & \begin{tabular}[c]{@{}l@{}}cohort b:   0.76±0.04\\    \\ cohort c: 0.80±0.015\end{tabular} \\ \hline
		Hsu et al\cite{RN71} & 2021 & Tumor & V-Net & \begin{tabular}[c]{@{}l@{}}Residual\\    \\ 3D\end{tabular} & T1C, CECT & 511 & 0.76 ± 0.03 \\ \hline
		\begin{tabular}[c]{@{}l@{}}Shirokikh\\    \\ et al\cite{RN72}\end{tabular} & 2021 & Tumor & U-Net & ROI & T1W & 1952 images & 0.64 ± 0.22 \\ \hline
		Lin et al\cite{RN73} & 2021 & Tumor & U-Net & BRATS & \begin{tabular}[c]{@{}l@{}}T1, T1C, T2W,\\    \\ T2-FLAIR\end{tabular} & 369 & \begin{tabular}[c]{@{}l@{}}WT: 0.92 ± 0.05\\    \\ TC: 0.89 ± 0.18\\    \\ ET: 0.85 ± 0.17\end{tabular} \\ \hline
		Huang et al\cite{RN74} & 2021 & Tumor & FCN & BRATS & T2W, T1C, T1, FLAIR & 384 & \begin{tabular}[c]{@{}l@{}}WT: 0.86\\    \\ TC:  0.73\\    \\ ET: 0.61\end{tabular} \\ \hline
		Ahmadi et al   \cite{RN44} & 2021 & Tumor & QAIS-DSNN & BRATS & \begin{tabular}[c]{@{}l@{}}T1, T1C, T2W,\\    \\ T2-FLAIR\end{tabular} & 145 & \begin{tabular}[c]{@{}l@{}}WT: 0.92\\    \\ ET: 0.75\\    \\ TC: 0.80\end{tabular} \\ \hline
		Lee et al\cite{RN62} & 2021 & Tumor & \begin{tabular}[c]{@{}l@{}}Dual Pathway\\    \\ U-Net\end{tabular} & 3D & 1.5 T T1C, T2W & 381 & 0.90±0.05 \\ \hline
		\begin{tabular}[c]{@{}l@{}}Eijgelaar\\    \\ et al\cite{RN75}\end{tabular} & 2020 & GBM & DeepMedic & BRATS,3D & T1W, T2W, T1C,   FLAIR & 751 & \begin{tabular}[c]{@{}l@{}}BRATS: 0.80\\    \\ Clinical: 0.49\\    \\ Sparely Labeled: 0.67\end{tabular} \\ \hline
		\begin{tabular}[c]{@{}l@{}}Rahmat\\    \\ et al\cite{RN75}\end{tabular} & 2020 & GBM & Deep Medic & 3D & \begin{tabular}[c]{@{}l@{}}3.0 T DTI,\\    \\ T2-FLAIR, T1C\end{tabular} & 80 & 0.82 ± 0.17 \\ \hline
		Ermis et al\cite{RN76} & 2020 & GBM & DenseNet & \begin{tabular}[c]{@{}l@{}}BRATS\\    \\ Dense\end{tabular} & \begin{tabular}[c]{@{}l@{}}T1W, T1C,\\    \\ T2-FLAIR, T2W\end{tabular} & 30 & \begin{tabular}[c]{@{}l@{}}WT: 0.83\\    \\ TC: 0.81\\    \\ ET: 0.81\end{tabular} \\ \hline
		Tang et al\cite{RN76} & 2020 & Glioma & U-Net &  & \begin{tabular}[c]{@{}l@{}}CT, T1-FLAIR,\\    \\ T2-FLAIR, T2W, T1C\end{tabular} & 59 & 0.818 \\ \hline
		Haensch et al\cite{RN25} & 2020 & HC & \begin{tabular}[c]{@{}l@{}}One Hundred\\    \\  Layers\\    \\ Tiramisu\end{tabular} & \begin{tabular}[c]{@{}l@{}}Dense\\    \\ 2.5D\end{tabular} & T1W & 45 & 0.67 \\ \hline
		\begin{tabular}[c]{@{}l@{}}Mlynarski\\    \\ et al\cite{RN77}\end{tabular} & 2020 & Multi-Organ & U-Net &  & T1W & 44 & \begin{tabular}[c]{@{}l@{}}hippocampus: 0.88\\    \\ pituitary: 0.80\\    \\ brain: 0.99\end{tabular} \\ \hline
		Zhou et al\cite{RN78} & 2020 & Tumor & FCN & \begin{tabular}[c]{@{}l@{}}ROI\\    \\ 2.5D\end{tabular} & T1C & 934 & 0.81 ± 0.15 \\ \hline
		Bousabarah et al\cite{RN20} & 2020 & Tumor & U-Net &  & 3.0 T T1C, T2W, T2-FLAIR & 509 & 0.60 \\ \hline
		Xue et al\cite{RN80} & 2019 & Tumor & FCN &  & 3.0 T T1W & 1201 & 0.85±0.08 \\ \hline
		\begin{tabular}[c]{@{}l@{}}Charron\\    \\ et al\cite{RN81}\end{tabular} & 2018 & Tumor & DeepMedic & 3D & T1C, T2-FLAIR, T1W & 182 & 0.79 \\ \hline
		Liu et al\cite{RN16} & 2017 & Tumor & DeepMedic & \begin{tabular}[c]{@{}l@{}}BRATS\\    \\ 3D\end{tabular} & 3.0 T T1C & 240 & \begin{tabular}[c]{@{}l@{}}TC: 0.75 ± 0.07\\    \\ ET: 0.81 ± 0.04\end{tabular} \\ \hline
	\end{longtable}
}


	\noindent 
	\subsection{Head and Neck}
	
	The head and neck (HN) region contains many small structures, making high-resolution and high-contrast imaging of great importance. MRI is especially preferred over CT imaging for patients with amalgam dental fillings due to the metallic content that can cause intense streaking artifacts on CT.\cite{RN82} In addition, MRI is the standard of care for nasopharyngeal carcinoma (NPC), leading to significant research attention on auto-segmentation algorithms for HN MR images. Other research efforts include segmentation of oropharyngeal cancer, glands, and lymph nodes in the American Association of Physicists in Medicine (AAPM)’s RT-MAC challenge\cite{RN84}, as well as multi-organ segmentation.
	
	Notable efforts include the two-stage multi-channel Seqseg architecture for NPC segmentation.\cite{RN97} Seqseg uses reinforcement learning to refine the position of the bounding box, implements residual blocks, recurrent channel and region-wise attention, and a custom loss function that emphasizes segmentation of the edges of the tumor. Outierial et al\cite{RN91} improves the dice score by 0.10 with a two-stage approach compared to single-state 3D U-Net for oropharyngeal cancer segmentation. For multiparametric MRI (mp-MRI), Deng et al\cite{RN85} concludes that the union output from T1W and T2W sequences has similar performance to T1C MRI, suggesting that contrast may not be necessary for NPC segmentation. Similarly, Wahid et al\cite{RN86} found that T1W and T2W sequences significantly improve performance, but dynamic contrast enhanced MRI (DCE) and diffusion weighted imaging (DWI) have little effect. Interesting approaches to gland and lymph node segmentation came out of the AAPM’s RT-MAC challenge, with Kawahara et al’s\cite{RN87} 2.5D GAN and Korte et al\cite{RN88} employing a 2-stage architecture. The first stage segments the OARs in low resolution to create a bounding box, followed by U-Net segmenting the ROI in high resolution. Jiang et al segments the parotid glands using T2W MRI and unpaired CT images with ground truth contours. First, sMRI is generated from the CT volumes using a GAN. In the second step, U-Net generates probabilistic segmentation maps for both the sMRI and MRI based on the CT ground truth contours. These maps, along with sMRI and MRI data, are then input into the organ attention discriminator, which is designed to learn finer details during training, ultimately producing the final segmentations.\cite{RN279}
	
	\begin{table}[]
		\centering
		\caption{HN Segmentation Studies}
		\label{tab:my-table}
		\resizebox{\textwidth}{!}{%
			\begin{tabular}{llllllll}
				\hline
				Study & Year & Target & \begin{tabular}[c]{@{}l@{}}Network\\    \\ Architecture\end{tabular} & \begin{tabular}[c]{@{}l@{}}Network \\    \\ Features\end{tabular} & \begin{tabular}[c]{@{}l@{}}Imaging\\    \\ Modalities\end{tabular} & Patient Number & DSC \\ \hline
				Dai et al\cite{RN89} & 2022 & Multi-organ & MS R-CNN & \begin{tabular}[c]{@{}l@{}}Attention\\    \\ Residual ROI\\    \\ 3D\end{tabular} & 1.5T T1W & 60 & \begin{tabular}[c]{@{}l@{}}optic chiasm: 0.61 ± 0.14\\    \\ oral cavity: 0.92 ± 0.07\end{tabular} \\ \hline
				Tao et al\cite{RN97} & 2022 & NPC & SeqSeg & \begin{tabular}[c]{@{}l@{}}Attention \\    \\ Residual\\    \\ ROI\\    \\ Recurrent\end{tabular} & \begin{tabular}[c]{@{}l@{}}T1W, T2W,\\    \\ T1C\end{tabular} & 596 & 0.80 \\ \hline
				Deng et al\cite{RN85} & 2022 & NPC & \begin{tabular}[c]{@{}l@{}}DenseNet,\\    \\ V-Net\end{tabular} & \begin{tabular}[c]{@{}l@{}}Dense\\    \\ 3D\end{tabular} & \begin{tabular}[c]{@{}l@{}}3.0T T1W, T2W, T1C\\    \\  (separately)\end{tabular} & 4478 & \begin{tabular}[c]{@{}l@{}}T1: 0.77±0.07\\    \\ T2: 0.76±0.07\end{tabular} \\ \hline
				Zhang et al\cite{RN91} & 2022 & NPC & AttR2U-Net & Attention & T1C & 93 & 0.82 \\ \hline
				Outeiral et al\cite{RN92} & 2022 & \begin{tabular}[c]{@{}l@{}}Oropharyngeal\\    \\ Cancer\end{tabular} & U-Net & \begin{tabular}[c]{@{}l@{}}ROI\\    \\ 3D\end{tabular} & T1W, T2W & 230 & 0.64 \\ \hline
				Jiang et al\cite{RN279} & 2022 & Paratoid glands & \begin{tabular}[c]{@{}l@{}}GAN,\\    \\ U-Net\end{tabular} & \begin{tabular}[c]{@{}l@{}}Attention\\    \\ Residual\end{tabular} & 3.0T T2W, CT & 181 & \begin{tabular}[c]{@{}l@{}}right parotid gland: 0.81 ± 0.05\\    \\ left parotid gland: 0.82 ± 0.03\end{tabular} \\ \hline
				\begin{tabular}[c]{@{}l@{}}Kawahara\\    \\ et al\cite{RN87}\end{tabular} & 2022 & \begin{tabular}[c]{@{}l@{}}Paratoid glands, \\    \\ Submandibular glands, lymph nodes\end{tabular} & GAN & 2.5D & 1.5T T2W & 55 & \begin{tabular}[c]{@{}l@{}}right lymph node: 0.75\\    \\ right parotid gland: 0.85\end{tabular} \\ \hline
				Li et al\cite{RN93} & 2021 & NPC & DenseNet & Dense & T1W & 30 & 0.872 \\ \hline
				Wahid et al\cite{RN86} & 2021 & \begin{tabular}[c]{@{}l@{}}Oropharyngeal\\    \\ Cancer\end{tabular} & \begin{tabular}[c]{@{}l@{}}Residual\\    \\ U-Net\end{tabular} & \begin{tabular}[c]{@{}l@{}}Residual\\    \\ 3D\end{tabular} & \begin{tabular}[c]{@{}l@{}}1.5 T T1W,\\    \\ T2W, DCE, DWI\end{tabular} & 30 & \begin{tabular}[c]{@{}l@{}}ALL MR Sequences:\\    \\ 0.71 ± 0.12\\    \\ T1W + T2W:\\    \\ 0.73 ± 0.12\end{tabular} \\ \hline
				Outeiral et al\cite{RN94} & 2021 & \begin{tabular}[c]{@{}l@{}}Oropharyngeal\\    \\ Cancer\end{tabular} & U-Net & 3D & T1W, T2W, T1C & 171 & 0.74 \\ \hline
				Korte et al\cite{RN88} & 2021 & \begin{tabular}[c]{@{}l@{}}Paratoid glands,\\    \\  Submandibular glands, lymph nodes\end{tabular} & U-Net & \begin{tabular}[c]{@{}l@{}}ROI\\    \\ 3D\end{tabular} & 1.5T T2W & 41 & \begin{tabular}[c]{@{}l@{}}LN Lvl IIIL:\\    \\ 0.56 ± 0.10\\    \\ Left Parotid:\\    \\ 0.86 ± 0.07\end{tabular} \\ \hline
				Ren et al\cite{RN95} & 2021 & Tumors & U-Net &  & PET, CT, T2W, T1W & 153 & 0.87 \\ \hline
				\begin{tabular}[c]{@{}l@{}}Gurney-Champion\\    \\ et al\cite{RN96}\end{tabular} & 2020 & Lymph nodes & U-Net & 3D & 1.5 T DWI & 48 & 0.87 \\ \hline
				Ke et al\cite{RN97} & 2020 & NPC & DenseNet & \begin{tabular}[c]{@{}l@{}}Dense\\    \\ 3D\end{tabular} & 3.0 T T1W & 4100 & 0.77 ± 0.07 \\ \hline
				Lin et al\cite{RN98} & 2019 & NPC & VoxResNet & \begin{tabular}[c]{@{}l@{}}Residual\\    \\ 3D\end{tabular} & T1W, T2W, T1C, T1-Fat Suppressed & 203 & 0.79 \\ \hline
			\end{tabular}%
		}
	\end{table}

	\noindent 
	\subsection{Abdomen, Heart, And Lung}

	In contrast to the brain, the abdomen is susceptible to respiratory and digestive motion of the patient often leading to poorly defined boundaries. While motion management techniques like patient breath-hold and not eating or drinking before treatment can mitigate these effects, the long acquisition time of MRI will inevitably lead to errors. Often physicians must rely on anatomical knowledge to deduce the boundaries of OARs. This makes segmentation challenging for CNN-based architectures which build from local context. In addition, registration errors make including multiple sequences impractical. OARs segmented in the abdomen include the liver, kidneys, stomach, bowel, and duodenum. The liver and kidneys are not associated with digestion and are relatively stable while the stomach, bowel, and duodenum are considered unstable. The duodenum is the most difficult for segmentation algorithms due to its small size, low contrast, and variability in shape. In addition, radiation induced duodenal toxicity is often dose-limiting in dose escalation studies making accurate segmentation of high importance.\cite{RN98} Similar problems occur in the heart and lung because of their periodic motion with the lung being particularly challenging since it is filled with low-signal air. However, MR segmentation of cardiac subregions have shown growing interest as these are not visible on CT and have different tolerances to radiation.\cite{RN99}
	
	The results are summarized in Table 3. Due to the large number of organs segmented in several of these studies, only the stomach and duodenum dice scores are reported to establish how the algorithms handle unstable organs. Zhang et al\cite{RN100} generates a composite image from the current slice, prior slice, and contour map to pre-dict the current segmentation with U-Net. Luximon et al\cite{RN100} takes a similar approach by having a phy-sician contour every 8th slice. These contours are then linearly interpolated and improved upon with a 2D Dense U-Net. The remaining studies do not require previous information and struggle to segment the duodenum. Ding et al\cite{RN101} improves upon a physician-defined acceptable contour rate by up to 39\% with an active contour model. A 3D Dense U-Net with sequential refinement networks is included in Fu et al\cite{RN102}. Morris et al segments heart substructures with a 2 channel 3D U-Net.\cite{RN103} Wang et al segments lung tumors with high accuracy relying on segmentation maps from previous weeks with the aim of adaptive radiation therapy (ART).\cite{RN104} An addition study by the same group feeds the features from the CNN into a GRU based RNN to predict tumor position over the next 3 weeks. Attention is included to weigh the importance of the prior weeks’ segmentation maps.\cite{RN105}
	
	\begin{table}[]
		\centering
		\caption{Abdomen, Heart, and Lung Segmentation Studies}
		\label{tab:my-table}
		\resizebox{\textwidth}{!}{%
			\begin{tabular}{llllllll}
				\hline
				Study & Year & Target & \begin{tabular}[c]{@{}l@{}}Network\\    \\ Architecture\end{tabular} & \begin{tabular}[c]{@{}l@{}}Network\\    \\ Features\end{tabular} & \begin{tabular}[c]{@{}l@{}}Imaging\\    \\ Modalities\end{tabular} & Patient Number & DSC \\ \hline
				\begin{tabular}[c]{@{}l@{}}Zhang\\    \\ et al\cite{RN91}\end{tabular} & 2022 & Multi-Organ & U-Net &  & 3T, T2w HASTE & 75 & \begin{tabular}[c]{@{}l@{}}DD: 0.88 ± 0.03\\    \\ Stomach: 0.92 ± 0.02\end{tabular} \\ \hline
				\begin{tabular}[c]{@{}l@{}}Ding\\    \\ et al\cite{RN101}\end{tabular} & 2022 & Multi-Organ & \begin{tabular}[c]{@{}l@{}}ResU-Net,\\    \\ Active Contour   Model\end{tabular} & \begin{tabular}[c]{@{}l@{}}Residual\\    \\ 3D\end{tabular} & 3T, T2w HASTE & 71 & \begin{tabular}[c]{@{}l@{}}DD: 0.49-0.69\\    \\ Stomach: 0.56-0.77\end{tabular} \\ \hline
				\begin{tabular}[c]{@{}l@{}}Luximon\\    \\ et al\cite{RN3}\end{tabular} & 2021 & Bowel Stomach & Dense U-Net & Dense & .35 T MRI & 116 & \begin{tabular}[c]{@{}l@{}}Bowel: 0.90 ± 0.04\\    \\ Stomach: 0.91 ±0.02\end{tabular} \\ \hline
				Morris et al\cite{RN103} & 2020 & \begin{tabular}[c]{@{}l@{}}Heart\\    \\ substructures\end{tabular} & U-Net & 3D & CT, T2W & 32 & \begin{tabular}[c]{@{}l@{}}Chambers:\\    \\ 0.88 ± 0.03\\    \\ Great Vessels:\\    \\ 0.85 ± 0.03\\    \\ pulmonary veins:\\    \\  0.77 ± 0.04\end{tabular} \\ \hline
				Chen et al\cite{RN98} & 2020 & Multi-Organ & Dense U-Net & \begin{tabular}[c]{@{}l@{}}Dense\\    \\ 2.5D\end{tabular} & 3.0 T T1W VIBE & 102 & \begin{tabular}[c]{@{}l@{}}DD:  0.80 ± 0.07\\    \\ Stomach: 0.92 ±   0.02\end{tabular} \\ \hline
				Wang et al\cite{RN104} & 2019 & Lung Cancer & FCN &  & \begin{tabular}[c]{@{}l@{}}3T T2W,\\    \\ prior contours\end{tabular} & 9 & 0.82 ± 0.10 \\ \hline
				Wang et al\cite{RN105} & 2019 & Lung Cancer & CNN, GRU & Recurrent Attention & 3T T2W & 10 & \begin{tabular}[c]{@{}l@{}}Week 4: 0.78 ± 0.22\\    \\ Week 5: 0.69 ± 0.24\\    \\ Week 6: 0.69 ± 0.26\end{tabular} \\ \hline
				Fu et al\cite{RN98} & 2018 & Multi-organ & Dense U-Net & \begin{tabular}[c]{@{}l@{}}Dense\\    \\ 3D\end{tabular} & .35 T MRI & 120 & \begin{tabular}[c]{@{}l@{}}DD: 0.66 ± 0.09\\    \\ Stomach: 0.85 ± 0.04\end{tabular} \\ \hline
			\end{tabular}%
		}
	\end{table}

	\noindent 
	\subsection{Pelvis}
	The anatomy of the pelvis allows both external beam radiation therapy (EBRT) and brachytherapy approaches for radiation therapy. Therefore, MRI segmentation studies have proposed methods to contour fiducial markers and catheters for cervical and prostate therapy, as well as tumors and OARs. However, a current challenge is that fiducials and catheters are designed for CT and are not optimal for MRI segmentation. For example, in prostate EBRT, gold fiducial markers localize the prostate with high contrast and correct for motion. However, metal does not emit a strong signal on MRI, so fiducials on MRI are characterized by an absence of signal, which can be confused with calcifications. Despite this, MRI is enabling treatments with higher tumor conformality. For instance, the gross tumor volume (GTV) of prostate cancer is not well delineated on CT but is often visible on MRI. In addition, the prostate apex is significantly clearer on MRI.\cite{RN106} MRI-based focal boost radiation therapy, in addition to a single dose level to the whole prostate, escalates additional dose to the GTV to reduce tumor recurrence.\cite{RN107}
	
	Table 4 shows relevant auto-segmentation techniques applied to the pelvic region. Shaaer et al\cite{RN108} segments catheters with a T1W and T2W MRI-based U-Net model and takes advantage of catheter continuity to refine the contours in post processing. Zabihollahy et al\cite{RN109} creates an uncertainty map of cervical tumors by retraining the U-Net model with a randomly set dropout layer. This technique is called Monte Carlo Dropout (MCDO). Cao et al\cite{RN23} takes pre-implant MRI and post-implant CT as input channels to their network. After preforming intra-observer variability analysis, they achieve performance more similar to a specialist radiation oncologist for cervical tumors in brachytherapy than a non-specialist. Eidex et al\cite{RN27} segments dominant intraprostatic lesions (DILs) and the prostate for focal boost radiation therapy with a Mask R-CNN based architecture. Sensitivity is found to be an important factor in evaluating model performance because weak models can appear strong by missing difficult lesions entirely. Figure 3 shows an example of automatic contours of the prostate and DIL on T2w MRI which would not be visible on CT. STRAINet\cite{RN110} realizes exceptional performance by utilizing a GAN with stochastic residual and atrous convolutions. In contrast with standard residual connections, each element of the input feature map which does not undergo convolution has a 1\% chance of being set to zero. Singhrao et al\cite{RN111} implements a pix2pix architecture for fiducial detection achieving 96\% detection with the misses caused by calcifications.
	
	\begin{figure}
		\centering
		\noindent \includegraphics*[width=6.50in, height=4.20in, keepaspectratio=true]{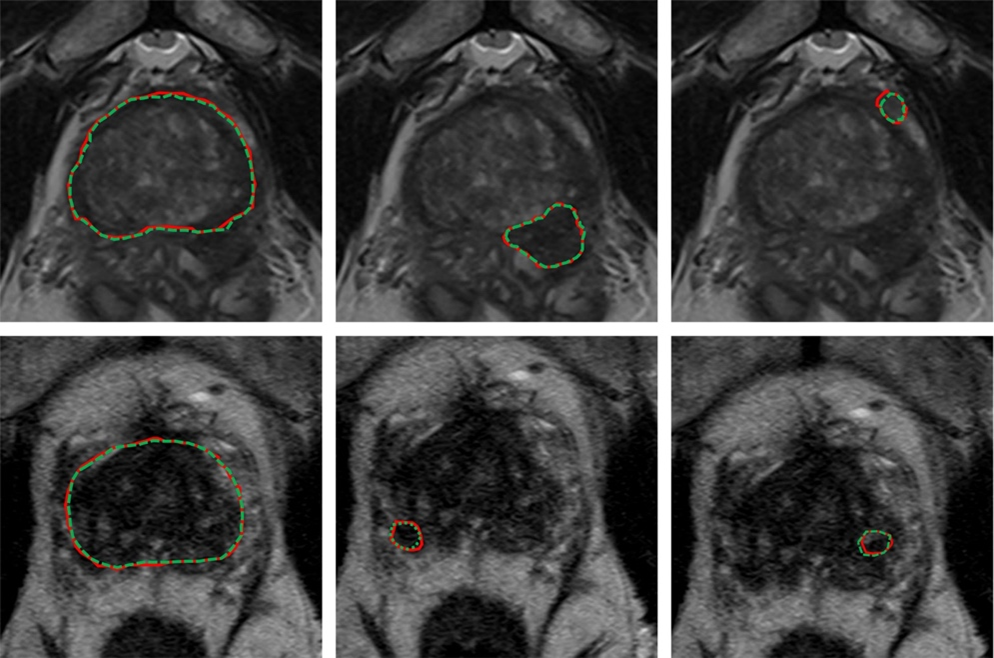}
		
		\noindent Figure 3. Expert (red) versus proposed auto-segmented (green dashed) prostate and DIL contours on axial MRI. From left to right: prostate manual and auto-segmented contours overlaid on MRI, and two DIL manual and auto-segmented contours overlaid on MRI. The upper and lower rows are representative of two patients. Reprinted by permission from John Wiley and Sons: Medical Physics, MRI-based prostate and dominant lesion segmentation using cascaded scoring convolutional neural network by Eidex et al\cite{RN27} © 2022. 
	\end{figure}
	
	\begin{table}[]
		\centering
		\caption{Pelvic Segmentation Studies}
		\label{tab:my-table}
		\resizebox{\textwidth}{!}{%
			\begin{tabular}{llllllll}
				\hline
				Study & Year & Target & \begin{tabular}[c]{@{}l@{}}Network\\    \\ Architecture\end{tabular} & \begin{tabular}[c]{@{}l@{}}Network\\    \\ Features\end{tabular} & \begin{tabular}[c]{@{}l@{}}Imaging\\    \\ Modalities\end{tabular} & \begin{tabular}[c]{@{}l@{}}Patient\\    \\ Number\end{tabular} & DSC \\ \hline
				Groendahl et al\cite{RN112} & 2022 & Anal Cancer & U-Net & ROI & \begin{tabular}[c]{@{}l@{}}3.0 T2W, DWI,   PET/CT,\\    \\ contrast CT\end{tabular} & 36 & \begin{tabular}[c]{@{}l@{}}PET, contrast CT:\\    \\ 0.83 ± 0.08\\    \\ PET, contrast CT,   T2W:\\    \\  0.81 ± 0.08\end{tabular} \\ \hline
				Shaaer et al\cite{RN108} & 2022 & \begin{tabular}[c]{@{}l@{}}Catheters \\    \\ (Cervix)\end{tabular} & U-Net &  & 1.5 T T1W, T2W & 20 & 0.59 ± 0.10 \\ \hline
				Zabihollahy et   al\cite{RN109} & 2022 & Cervical Cancer & U-Net & 3D & 1.5T T2W MRI & 123 & 0.85 ± 0.03 \\ \hline
				Cao et al\cite{RN113} & 2022 & Cervical Cancer & Dual-path CNN & Residual & T2W, CT & 65 & \begin{tabular}[c]{@{}l@{}}Small: 0.65 ± 0.03\\    \\ Medium: 0.79 ± 0.02\\    \\  Large: 0.75 ± 0.04\end{tabular} \\ \hline
				Yoganathan et al\cite{RN114} & 2022 & \begin{tabular}[c]{@{}l@{}}Cervical Cancer\\    \\  Multi-Organ\end{tabular} & \begin{tabular}[c]{@{}l@{}}ResNet50,\\    \\ InceptionResNetV2\end{tabular} & \begin{tabular}[c]{@{}l@{}}Residual\\    \\ 2.5D\end{tabular} & 1.5T T1W & 39 & GTV:   0.62 ± 0.14 \\ \hline
				Breto et al\cite{RN115} & 2022 & \begin{tabular}[c]{@{}l@{}}Cervical Cancer\\    \\ Multi-Organ\end{tabular} & Mask R-CNN & Residual & \begin{tabular}[c]{@{}l@{}}0.35T\\    \\ MRIdian\end{tabular} & 15 & GTV: 0.67 ± 0.30 \\ \hline
				Li et al\cite{RN276} & 2022 & \begin{tabular}[c]{@{}l@{}}Liver,\\    \\ Kidney,\\    \\ Cervical Cancer\end{tabular} & nnU-Net &  & T2W & 6 & \begin{tabular}[c]{@{}l@{}}Liver GTV:   0.94 ± 0.01\\    \\ Kidney GTV:   0.95 ± 0.02\\    \\ Cervix GTV:  0.97 ± 0.02\end{tabular} \\ \hline
				Fransson et al\cite{RN116} & 2022 & Prostate and OARs & U-Net & ROI & 3T, T2W & 17 & \begin{tabular}[c]{@{}l@{}}CTV: 0.92 ± 0.03\\    \\ Bladder: 0.93 ±   0.07 \\    \\ Rectum: 0.84 ± 0.10\end{tabular} \\ \hline
				Eidex et al\cite{RN21} & 2022 & Prostate Cancer & Mask R-CNN & \begin{tabular}[c]{@{}l@{}}Residual\\    \\ ROI\\    \\ 3D\end{tabular} & T1W & 77 & \begin{tabular}[c]{@{}l@{}}Prostate: 0.90 ± 0.09\\    \\ DIL: 0.84 ± 0.12\end{tabular} \\ \hline
				Li et al\cite{RN117} & 2021 & Anal Cancer & U-Net & Attention & Not Specified & 304 & 0.98 \\ \hline
				Huang et al\cite{RN118} & 2021 & \begin{tabular}[c]{@{}l@{}}Colorectal \\    \\ Cancer\end{tabular} & RU-Net & \begin{tabular}[c]{@{}l@{}}ROI\\    \\ 3D\end{tabular} & T2W & 64 & 0.76 \\ \hline
				Zabihollahy et   al\cite{RN119} & 2021 & \begin{tabular}[c]{@{}l@{}}Female Bladder,   Rectum,\\    \\ Sigmoid\\    \\ Colon\end{tabular} & \begin{tabular}[c]{@{}l@{}}3D U-Net\\    \\ 3D Dense U-Net\end{tabular} & \begin{tabular}[c]{@{}l@{}}Dense\\    \\ ROI\\    \\ 3D\end{tabular} & 1.5 T T2W & 129, 52 & \begin{tabular}[c]{@{}l@{}}Bladder:   0.94 ± 0.05\\    \\ Rectum: 0.88 ±   0.04 Sigmoid: 0.80 ± 0.05\end{tabular} \\ \hline
				Cha et al\cite{RN113} & 2021 & Prostate & DeepLabV3 + & Residual & 3.0 T T2W, sCT & 50 & \begin{tabular}[c]{@{}l@{}}Prostate: 0.89\\    \\ Bladder: 0.99\end{tabular} \\ \hline
				Comelli et al\cite{RN120} & 2021 & Prostate & E-Net &  & T1W & 85 & 0.91 \\ \hline
				Savenije et al\cite{RN121} & 2020 & \begin{tabular}[c]{@{}l@{}}Bladder, \\    \\ Rectum, Femur\end{tabular} & Deep Medic & 3D & 3.0 T T1W & 150 & \begin{tabular}[c]{@{}l@{}}Bladder: 0.96 ±   0.02\\    \\ Rectum: 0.88 ± 0.05\\    \\ femurs:  0.97 ±0.01\end{tabular} \\ \hline
				Dai et al\cite{RN122} & 2020 & \begin{tabular}[c]{@{}l@{}}Catheters\\    \\  (prostate)\end{tabular} & AGU-Net & \begin{tabular}[c]{@{}l@{}}Attention\\    \\ 3D\end{tabular} & 1.5 T T2W & 20 & \begin{tabular}[c]{@{}l@{}}Average   displacement:\\    \\ 0.37±1.68 mm\end{tabular} \\ \hline
				\begin{tabular}[c]{@{}l@{}}Singhrao\\    \\ et al\cite{RN111}\end{tabular} & 2020 & \begin{tabular}[c]{@{}l@{}}Fiducials \\    \\ (prostate)\end{tabular} & pix2pix (GAN) &  & T1W & 56 & 0.67 \\ \hline
				Gustafsson et al\cite{RN123} & 2020 & \begin{tabular}[c]{@{}l@{}}Fiducials\\    \\  (prostate)\end{tabular} & HighRes3DNet & \begin{tabular}[c]{@{}l@{}}Residual\\    \\ 3D\end{tabular} & 3T T2W & 326 & 0.98 ± 0.002 \\ \hline
				Sanders et al\cite{RN124} & 2020 & Prostate & DenseNet-201 &  & T1W, T2W, T1C & 200 & \begin{tabular}[c]{@{}l@{}}Prostate: 0.90 ±   0.04\\    \\  Bladder: 0.91 ± 0.06\\    \\  Rectum: 0.96 ± 0.04\end{tabular} \\ \hline
				\begin{tabular}[c]{@{}l@{}}da Silva\\    \\ et al\cite{RN125}\end{tabular} & 2020 & Prostate & \begin{tabular}[c]{@{}l@{}}Hybrid atlas,\\    \\ active contour\end{tabular} &  & T2W & 56 & 0.85 \\ \hline
				Chen et al\cite{RN126} & 2020 & Prostate cancer & MB-U-Net &  & 3.0 T T2W, ADC, DWI & 136 & 0.63 \\ \hline
				Zaffino et al\cite{RN127} & 2019 & \begin{tabular}[c]{@{}l@{}}Catheters\\    \\  (Cervix)\end{tabular} & U-Net & 3D & T2W & 50 & 0.60 ± 0.17 \\ \hline
				Yang et al\cite{RN128} & 2019 & Prostate & MICS-Net &  & T2W, CT & 22 & 0.83 ± 0.04 \\ \hline
				\begin{tabular}[c]{@{}l@{}}Elguindi\\    \\ et al\cite{RN129}\end{tabular} & 2019 & Prostate and OARs & Deep LabV3+ & Residual & T2W & 50 & \begin{tabular}[c]{@{}l@{}}CTV:   0.83 ± 0.06\\    \\ Bladder:   0.93 ± 0.04 \\    \\ Rectum:  0.82 ± 0.05\end{tabular} \\ \hline
				Nie et al\cite{RN110} & 2019 & Prostate and OARS & STRAINet (GAN) & Residual & 3.0 T T1W & 35 & \begin{tabular}[c]{@{}l@{}}Prostate: 0.91 ± 0.01   \\    \\ Bladder: 0.97   ± 0.01 \\    \\ Rectum: 0.91   ± 0.03\end{tabular} \\ \hline
				Feng et al\cite{RN130} & 2018 & Prostate and OARs & ResNet & Residual & Not specified & 40 & \begin{tabular}[c]{@{}l@{}}Prostate: 0.90 ± 0.02\\    \\  Bladder: 0.96 ± 0.01 \\    \\ Rectum: 0.89   ± 0.03\end{tabular} \\ \hline
				\begin{tabular}[c]{@{}l@{}}Wang\\    \\ et al\cite{RN26}\end{tabular} & 2018 & \begin{tabular}[c]{@{}l@{}}Rectal\\    \\ Cancer\end{tabular} & U-Net & 2.5D & 3.0T T2W & 93 & 0.74 ± 0.14 \\ \hline
			\end{tabular}%
		}
	\end{table}

	\bigbreak
	
	\noindent 
	\section{IMAGE SYNTHESIS}
	
	Synthesis is an exciting field of research, defined as translating one imaging modality into another. Benefits of synthesis include avoiding potential artifacts, reducing patient cost and discomfort, and avoiding radiation exposure.\cite{RN131} In addition, utilizing multiple modalities introduces registration errors which can be avoided with synthetic images. Current methods in MRgRT include synthesis of sCT from MRI, sMRI from CT, and relative proton stopping power images from MRI. Other areas of synthesis research include creating higher resolution MRI (super-resolution) and predicting organ displacement based on periodic motion in 4D MRI. Segmentation can also be thought of as a special case of synthesis because the input MRI is translated into voxel-wise masks which assume discrete values according to their class. The distinction between synthesis and segmentation is particularly muddied when the segmentation ground truth is from a different imaging modality.\cite{RN132}
	
	Synthesis architectures are fundamentally interchangeable with segmentation architectures but have diverged in practice. For example, U-Net, described in detail in Section 3, is the predominant backbone in both areas. However, synthesis models require that the entire image be translated, so that they do not include two-stage architectures and are dominated by generational adversarial network (GAN) -based architectures. The GAN is comprised of a CNN or self-attention-based generator which generates synthetic images. The generator competes with a discriminator which attempts to correctly classify synthetic and real images. As the GAN trains, a loss function is applied to the discriminator when it mislabels the image, whereas a loss function is applied to the generator when the discriminator is correct. The model is ideally considered trained once the discriminator can no longer correctly identify the synthetic images. Conditional GANs (cGANs) expand on the standard GAN by also inputting a vector with random values or additional information into both the generator and discriminator.\cite{RN133} In the case of MRI, the values of the vector can correspond to the MRI sequence type and clinical data to account for differences in patient population and setup. The CycleGAN adds an additional discriminator and generator loop.\cite{RN134} For example, an 
	
	MRI would be translated into a sCT. The sCT would then be translated into a sMRI. Since the input is ultimately tested against itself, this allows for training with unpaired data. The need for co-registration is eliminated but requires significantly more data to achieve comparable results with paired training.
	
	Despite their success, GANs can be unstable during training and struggle in difficult synthesis problems. One way to improve its performance is with the Wasserstein GAN (WGAN)\cite{RN135}. Instead of the discriminator classifying the images as real or fake, the WGAN measures the probability distributions of the real and fake images and finds the distance between them in the form of the Wasserstein distance. The discriminator attempts to maximize this distance while the generator attempts to minimize it. The WGAN approach often improves stability and performance. Although not limited to WGANs, spectral normalization is often included which constricts the training weights of the discriminator such that the gradient cannot explode. Another approach, claiming better performance than the WGAN, is the relativistic GAN (RGAN)\cite{RN136}. The RGAN claims that the generator should, in addition to increasing the probability that synthetic images appear realistic, increase the probability that real images appear fake to the discriminator. Without this condition, the discriminator will conclude that every image it comes across is real in the late stages of training with a well-trained generator. This goes against the priori knowledge that half of the images are fake. A standard GAN can be converted to a RGAN by modifying its loss function.
	
	\noindent 
	\subsection{MRI-Based Synthetic CT}
	
	MRI-based sCT is the most extensively researched and influential application of synthesis models in radiation therapy. While MR images provide excellent soft tissue contrast, they do not contain the necessary attenuation information for dose calculation that is embedded in CT images. Owing to this limitation, CT has traditionally been the workhorse for treatment planning while MRI has been relegated to diagnostic applications. However, CT suffers from lower soft tissue contrast and imparts a non-negligible radiation dose, especially for patients receiving standard fractionated image guided radiation therapy (IGRT). In addition, metallic materials found in dental work and implants can lead to severe artifacts in CT, reducing the quality of the treatment plan. By augmenting CT with sCT, these problems can be avoided. Furthermore, according to the “As Low As Reasonably Achievable” (ALARA) principle, the replacement of CT with sCT for an MRI only workflow could be justified with its high accuracy, especially in radiosensitive populations like pediatric patients.\cite{RN137, RN138}
	
	Calculation of dose distribution using MRI-based sCT can be enhanced by replacing traditional Monte Carlo simulation (MC) techniques with deep learning. MC accurately predicts the dose distribution based on physical principles, including the electron return effect (ERE), which adds additional dose to boundaries with different proton densities in the presence of a magnetic field. However, the technique can be extremely slow, as it relies on randomly generating paths of tens of thousands of particles. The higher number of particles reduces dosimetric uncertainty. This problem is particularly noticeable in proton therapy, where MC or pencil beam algorithm (PBA) calculations can take several minutes on a CPU, and it can take hours to optimize a single treatment plan.\cite{RN139} As a result, compromises must be made in clinical practice between dosimetric uncertainty, MC run time, and treatment plan optimization. Deep learning methods show exceptional potential to improve upon MC dose calculation models. Once trained, deep learning algorithms take only a few seconds to synthesize a dose distribution. In addition, they can be trained on extremely high accuracy MC generated dose distributions that would be impractical in everyday clinical practice. 
	
	The primary challenge to sCT methods is the accurate reconstruction of bone and air, due to their low proton density and weak signal. This can make it difficult for sCT to distinguish between the two, leading to large errors. In addition, further complicating the issue is that bone makes up a small fraction of the patient volume in radiation therapy tasks or applications which is similar to the “small tumor problem” seen in segmentation. Other issues that can arise include small training sets, misalignment between CT and MRI, and causes of high imaging variability such as intestinal gas. 
	
	To evaluate sCT performance, various metrics are used to compare voxel values between the ground truth CT and sCT. The most common metric is the mean absolute error (MAE)\cite{RN141, RN140} which is reported in tables 5 and 6 if available. The MAE is defined below in Eq 2, where xi and yi are the corresponding voxel values of the CT and sCT, respectively, and n is the number of voxels.
	
	\begin{equation} 
		MAE=\Sigma_{i=1}^n|y_i– x_i|/n   
	\end{equation}
	
	The MAE is typically reported in Hounsfield units (HU) but can also be dimensionless if reported with normalized units. Other common metrics in literature are the mean error\cite{RN142}, which forgoes the absolute value in MAE, the mean squared error (MSE)\cite{RN143}, which substitutes absolute value for the square, and the Structural Similarity Index (SSIM), which varies from -1 to 1 where -1 represents extremely dissimilar images and 1 reperesents identical images.\cite{RN144} A full discussion of these metrics can be found in Necasova et al.\cite{RN145} Since sCT is primarily intended for treatment planning, dosimetric quantities which measure the deviation between CT- and sCT-derived plans are often reported. One of the most common metrics is gamma analysis. Repurposed as a metric to compare treatment plan dose to actual dose on LINACs, gamma analysis looks at each point on the dose distribution and evaluates if the acceptance criteria are met. The American Association of Physicists in Medicine (AAPM) Task Group 119 recommends a low dose threshold of 10\%, meaning that points which receive less than 10\% of the maximum dose are excluded from the calculation. Other metrics include the mean dose difference and the minimum dose delivered to 95\% of the clinical treatment volume (D95) difference.
	
	Sampling notable MRI-based sCT works for photon radiation therapy, several take advantage of cGANs to include additional information. Liu et al improves upon the CycleGAN by including a dense block, which captures structural and textural information and better handles local mismatc¬hes between MRI and ground truth CT images. In addition, a compound loss function with adversarial and distance losses improves boundary sharpness. An example patient is shown in Figure 4.\cite{RN274}  Hsu et al proposes a 2.5D method by training a pix2pix-based architecture with axial, sagittal, and coronal MRI slices.\cite{RN146} A conditional CycleGAN in Boni et al passes in MR manufacturer information and achieves good results despite using unpaired data and different centers for their training and test sets.\cite{RN147} Many studies also experiment with multiple sequences. Massa et al trains a U-Net with Inception-V3 blocks on 1.5T T1W, T2W, T1C, and FLAIR sequences separately and finds no statistical difference.\cite{RN148} However, Koike et al uses multiple MR sequences for sCT generation employing a cGAN to provide better image quality and dose distribution results compared with those from only a single T1W sequence.\cite{RN149} Dinkla et al finds that sCT removes dental artifacts.\cite{RN150} Farjam et al implements a custom loss function, which reweights contributions from bone.\cite{RN151} Wang et al uses a 3D pix2pix architecture, averaging results from 4x4x4, 6x6x6, and 8x8x8 filters to improve their results.\cite{RN152} Instead of using a GAN, Li et al simulates one with a fixed discriminator by pretraining a VGG16 loss network to discriminate between sCT and CT\cite{RN153}. Reaungamornrat et al decomposes features into modality specific and modality invariant spaces between high- and low-resolution Dixon MRI with the Huber distance.\cite{RN154} In addition, separable convolutions are used to reduce parameters, and a relativistic loss function is applied to improve training stability. Finally, Zhao et al represents the first MRI-based sCT paper to implement a hybrid transformer-CNN architecture outperforming other state-of-the-art methods. Their method implements a conditional GAN. The generator consists of CNN blocks in the shallow layers to capture local context and save computational resources, while transformers are used in deeper layers to provide better global context.\cite{RN155}
	
	\begin{figure}
		\centering
		\noindent \includegraphics*[width=6.50in, height=4.20in, keepaspectratio=true]{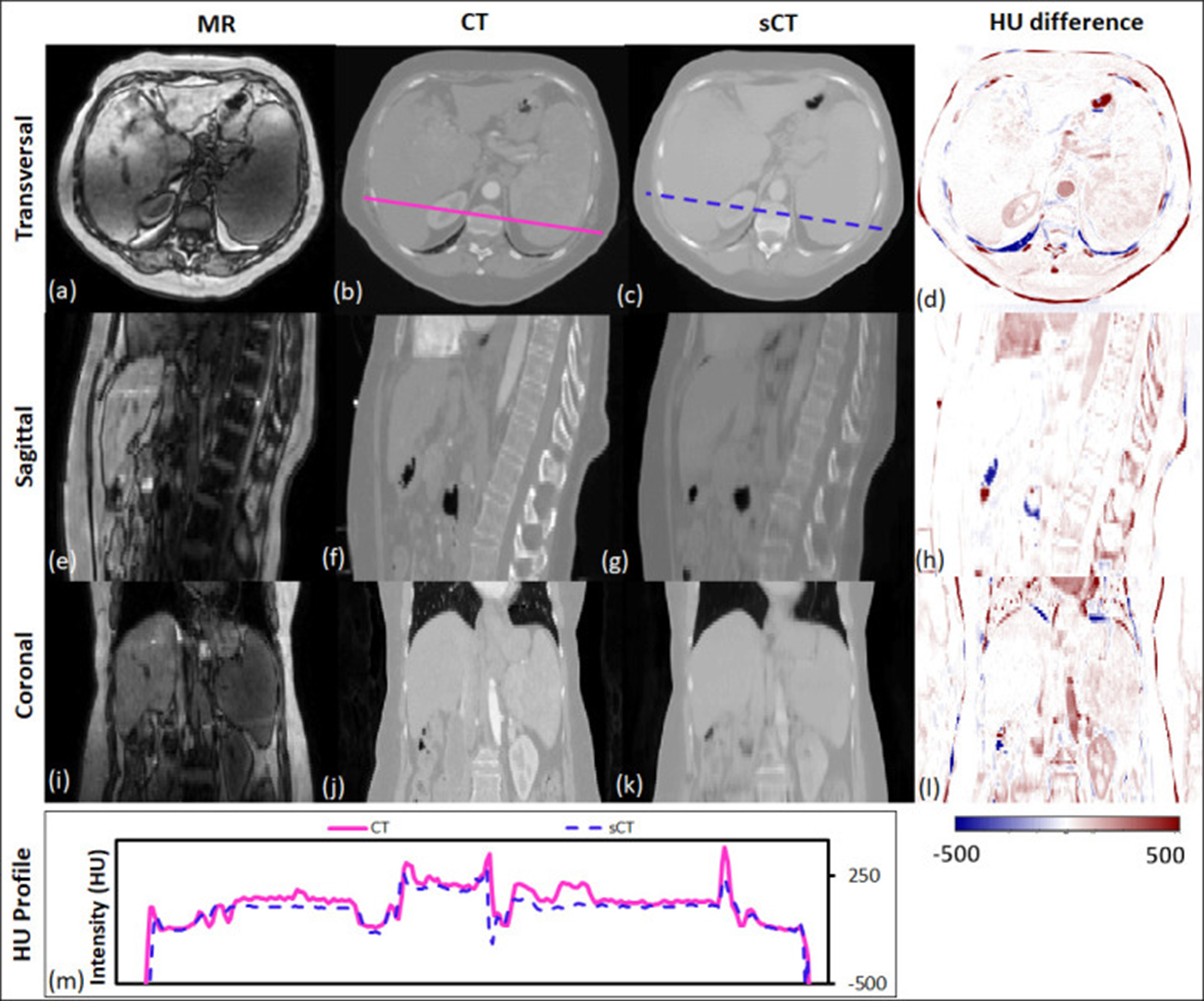}
		
		\noindent Figure 4. Traverse, sagittal, and coronal images of a representative patient. MRI, CT and sCT images and the HU difference map between CT and sCT are presented. The CT (solid line) and sCT (dashed line) voxel-based HU profiles of the traverse images are compared in the lowermost panel. Reprinted by permission from British Journal of Radiology, MRI-based treatment planning for liver stereotactic body radiotherapy: validation of a deep learning-based synthetic CT generation method by Liu et al.\cite{RN274}© 2019. 
	\end{figure}
	
	Generating sCTs from MRI for the purposes of proton therapy is not fundamentally different from the process for photon therapy. However, proton therapy takes advantage of the Bragg peak, which concentrates the radiation in a small region to spare healthy tissue. While this is beneficial, this puts a tighter constraint on sCT errors. Another difference is that sCT images must first be converted to relative proton stopping power maps before they can be used in treatment planning. Therefore, directly generating synthetic proton relative stopping power (sRPSP) maps instead of sCT would be ideal. Boron therapy is a form of targeted radiation therapy in which boronated compounds are delivered to the site of the tumor and irradiated with neutrons. The boron undergoes a fission reaction, releasing alpha particles that kill the tumor cells. However, the targeting mechanism typically relies on targeting cancer cells’ high metabolic rate. Epidermal tissue that also has a high metabolic rate uptakes boron, making skin dose an important concern in boron therapy. Therefore, methods for generating sCT images for boron therapy should emphasize accurate reconstruction around the skin. Shown in Table 6, many methods show high dosimetric accuracy for proton therapy. Liu et al develops a conditional cycleGAN to synthesize both high and lower energy CT. Multiple loss functions are also used to accurately classify and recreate the sCT.\cite{RN156} Wang et al creates the first synthetic relative proton stopping power maps from MRI with a cycleGAN and loss function to take advantage of paired data. Their method achieves an excellent MAE of 42 ± 13 HU, but struggles with dosimetric accuracy.\cite{RN6} Maspero et al achieves a 2\%/2mm gamma pass rate above 99\% for proton therapy by averaging predictions from three separate GANs trained on axial, sagittal, and coronal views, respectively.\cite{RN157} Replacing traditional MC dose calculation methods, Tsekas et al generates VMAT (volumetric modulated arc therapy) dose distributions in static positions with sCT. Additionally, parameters include a mask of the tissue exposed to the beam, the distance from LINAC source, the distance from central beam, and the radiological depth.\cite{RN158} These parameters are input into a 3D U-Net, significantly increasing processing speed.  Finally, SARU, a self-attention Res-UNet, lowers skin dose for boron therapy, achieving better results than the pix2pix method.\cite{RN159}
	
		\begin{landscape}
		{\small\tabcolsep=3pt  
		\begin{longtable}{ >{\bfseries\footnotesize}llllllllll}
			\caption{MRI-based Synthetic CT Studies for Photon Therapy}
			\label{tab:my-table}\\
			\hline
			Study & Year & Site & \begin{tabular}[c]{@{}l@{}}Network\\    \\ Architecture\end{tabular} & \begin{tabular}[c]{@{}l@{}}Network\\    \\ Features\end{tabular} & GAN Type & \begin{tabular}[c]{@{}l@{}}Imaging\\    \\ Modalities\end{tabular} & \begin{tabular}[c]{@{}l@{}}Patient\\    \\ Number\end{tabular} & MAE & Dosimetry \\ \hline
			\endfirsthead
			\endhead
			\begin{tabular}[c]{@{}l@{}}Ranjan\\    \\ et al\cite{RN160}\end{tabular} & 2022 & Brain & U-Net &  & pix2pix & T2W & 18 & 0.03 ± 0.02 & N/A \\ \hline
			Wang et al\cite{RN152} & 2022 & Brain & U-Net & 3D & pix2pix & 1.5T T1W & 31 & \begin{tabular}[c]{@{}l@{}}MSE:\\    \\ 0.12 ± 0.04 \%\end{tabular} & N/A \\ \hline
			\begin{tabular}[c]{@{}l@{}}Jabbarpour\\    \\ et al\cite{RN161}\end{tabular} & 2022 & Brain & FCN & Residual & cycleGAN & 3T T1W, T2W   (separately) & 189 & 61.9 ± 22.6 HU & \begin{tabular}[c]{@{}l@{}}2\%/2 mm:\\    \\ 95.0 ± 3.7\%\end{tabular} \\ \hline
			Scholey et al\cite{RN162} & 2022 & H\&N & U-Net & 3D & N/A & 1.5T T1W & 120 & \begin{tabular}[c]{@{}l@{}}whole body:\\    \\ 93.3 ± 27.5\\    \\ soft tissue:\\    \\ 78.2 ± 27.5\\    \\ bone:\\    \\ 138.0 ± 43.4\\    \\ (HU)\end{tabular} & 2\%/2 mm:   96.8 ± 2.6\% \\ \hline
			Florkow et al\cite{RN163} & 2022 & \begin{tabular}[c]{@{}l@{}}Hip\\    \\ Pelvis\end{tabular} & U-Net & 3D & N/A & 3.0 T T1W & 30 & \begin{tabular}[c]{@{}l@{}}Femur: 23±24\\    \\ Pelvis:\\    \\ -15±29\\    \\ (Mean Error HU)\end{tabular} & N/A \\ \hline
			Li et al\cite{RN164} & 2022 & Liver & U-Net &  & N/A & .35T sim & 37 & 35.6 HU & N/A \\ \hline
			Lenkowicz et al\cite{RN165} & 2022 & Lung & U-Net &  & pix2pix & .35T sim & 60 & 54.9 ± 10.5   HU & \begin{tabular}[c]{@{}l@{}}2\%/2mm:\\    \\ 96.1 ± 5.1\%\end{tabular} \\ \hline
			\begin{tabular}[c]{@{}l@{}}Reaunga-mornrat\\    \\ et al\cite{RN154}\end{tabular} & 2022 & Pelvis & FCN & Residual & relativistic GAN & \begin{tabular}[c]{@{}l@{}}High and low res\\    \\ Dixon MRI\end{tabular} & 45 & \begin{tabular}[c]{@{}l@{}}Median\\    \\ normalized mutual\\    \\ information:\\    \\ 1.28\end{tabular} & N/A \\ \hline
			O’Connor et al\cite{RN166} & 2022 & Pelvis & U-Net &  & cGAN & 4.0 T T1 VIBE Dixon & 40 & \begin{tabular}[c]{@{}l@{}}whole body:\\    \\ 34.7 ± 5.1\\    \\ bone:\\    \\ 109.4 ± 12.3\\    \\ soft tissue:\\    \\ 25.2 ± 3.4 (HU)\end{tabular} & 3\%/2 mm: 99.8\% \\ \hline
			Tsekas et al\cite{RN158} & 2022 & \begin{tabular}[c]{@{}l@{}}Abdomen\\    \\ Torso\end{tabular} & U-Net & 3D & N/A & 1.5 T1W & 124 & \begin{tabular}[c]{@{}l@{}}speed:\\    \\ 1.5 seconds\\    \\ per segment\end{tabular} & \begin{tabular}[c]{@{}l@{}}2\% / 2 mm:\\    \\ 96.3\% ± 4.2\%\\    \\ (for deep learning   dose calculation)\end{tabular} \\ \hline
			Zhao et al\cite{RN155} & 2022 & Pelvis & Transformer-CNN   hybrid & Residual   Transformer & cGAN & T2W & 19 & 45.1 HU & N/A \\ \hline
			Hsu et al\cite{RN146} & 2022 & Prostate & U-Net & 2.5D & pix2pix & .35T sim & 57 & \begin{tabular}[c]{@{}l@{}}pelvis:\\    \\ 30.1 ± 4.2\\    \\ soft tissue:\\    \\ 19.6 ± 2.3\\    \\ bone:\\    \\ 158.5 ± 26.0 (HU)\end{tabular} & 2\%/2 mm: 99.9\% \\ \hline
			Olber et al\cite{RN167} & 2021 & Abdomen & Dense U-Net & Dense & GAN & .35T sim & 89 & \begin{tabular}[c]{@{}l@{}}No gas:\\    \\ 90 ± 29 HU\\    \\ Gas:\\    \\ 143 ± 29 HU\end{tabular} & \begin{tabular}[c]{@{}l@{}}3\%/3 mm:\\    \\ well matched:\\    \\ 98.3 ± 1.3\%,\\    \\ poorly matched:   93.9 ± 9.8\%\end{tabular} \\ \hline
			Kang et al\cite{RN168} & 2021 & \begin{tabular}[c]{@{}l@{}}Abdomen\\    \\ Pelvis\\    \\ Thorax\end{tabular} & U-Net & \begin{tabular}[c]{@{}l@{}}Residual\\    \\ 2.5D\end{tabular} & cycleGAN & 0.35T sim & 90 & 59.2±5.8 HU & \begin{tabular}[c]{@{}l@{}}2\%/ 2mm,\\    \\ 10\% LDT: \textgreater{}97\%\end{tabular} \\ \hline
			Lerner et al\cite{RN169} & 2021 & Brain & FCN & 3D & N/A & Dixon MRI & 20 & Body: 62.2±4.1   Brain: 9.5±0.7 Bone: 173.8±18.2 & 2\%/ 2 mm:   99.8±0.2 \\ \hline
			Yuan et al\cite{RN143} & 2021 & Brain & Res U-Net & Residual & N/A & 1.5T T1W & 30 & 86.6±34.1 HU & D95 Difference:   1.1\% \\ \hline
			Liu et al\cite{RN170} & 2021 & Brain & ResNet & Residual & GAN & T1W & 12 & N/A & \begin{tabular}[c]{@{}l@{}}2\%/2mm:\\    \\ 99.9 ± 0.2\%\end{tabular} \\ \hline
			Koerkamp et al\cite{RN171} & 2021 & Breast & \begin{tabular}[c]{@{}l@{}}Not\\    \\  Specified\end{tabular} &  & revGAN & 1.5T T1W & 39 & 106 HU & \begin{tabular}[c]{@{}l@{}}2\%/2 mm,\\    \\ 10\% LDT: 99.4\%\end{tabular} \\ \hline
			Baydoun et al\cite{RN172} & 2021 & Cervix & U-Net &  & cGAN & T2W & 11 & 115.74 ± 21.84 HU & N/A \\ \hline
			Olin et al\cite{RN173} & 2021 & H\&N & U-Net & 3D & N/A & Dixon MRI & 6,17 & \begin{tabular}[c]{@{}l@{}}External:\\    \\ 78 ± 13 HU\\    \\ Local:\\    \\ 76 ± 12 HU\end{tabular} & \begin{tabular}[c]{@{}l@{}}2\%/ 2mm:\\    \\ 98.8 ± 0.8\%\end{tabular} \\ \hline
			Liu et al\cite{RN174} & 2021 & H\&N & Z-Net, FCN &  & CycleGAN & Dixon MRI & 164 & 0.04 & N/A \\ \hline
			Touati et al\cite{RN175} & 2021 & H\&N & U-Net &  & CycleGAN & 3.0T T1W & 56 & 45.3 ± 1.9 HU & N/A \\ \hline
			Song et al\cite{RN176} & 2021 & NPC & U-Net &  & N/A & 1.5T T1W & 35 & 125.6 HU & \begin{tabular}[c]{@{}l@{}}3\%/2mm:\\    \\ 97.7 ± 0.7\%\end{tabular} \\ \hline
			Ma et al\cite{RN177} & 2021 & NPC & U-Net &  & pix2pix & 3.0T T1W & 20 & 102.6 ± 11.4 HU & \begin{tabular}[c]{@{}l@{}}2 mm/3\%,\\    \\ 10\% LDT:\\    \\ 99.1\% ± 0.3\%\end{tabular} \\ \hline
			Szalkowski et al\cite{RN178} & 2021 & Pelvis & FCN & 3D & GAN & T2W & 11 & 72.9 ± 88.1 HU & N/A \\ \hline
			Boni et al\cite{RN147} & 2021 & Pelvis & \begin{tabular}[c]{@{}l@{}}Not \\    \\ Specified\end{tabular} &  & \begin{tabular}[c]{@{}l@{}}conditional\\    \\ CycleGAN\end{tabular} & 1.5T, 3.0T T2W & 38 & 59.8 HU & \begin{tabular}[c]{@{}l@{}}2\%/ 2mm:\\    \\ 95.5 ± 2.2\%\end{tabular} \\ \hline
			Bird et al\cite{RN179} & 2021 & \begin{tabular}[c]{@{}l@{}}Pelvis\\    \\ (ano-rectal)\end{tabular} & U-Net &  & pix2pix & 1.5T T2W & 90 & 35.1 ± 7.9 HU & \begin{tabular}[c]{@{}l@{}}2\%/ 2mm:\\    \\ 99.8 ± 0.1\%\end{tabular} \\ \hline
			Yoo et al\cite{RN180} & 2021 & Prostate & FCN & Residual & cycleGAN & T2W & 113 & 96.95 ± 10.32 HU & \begin{tabular}[c]{@{}l@{}}2\%/ 2mm:\\    \\ 93.9 ± 3.2\%\end{tabular} \\ \hline
			Farjam et al\cite{RN151} & 2021 & Prostate & U-Net &  & N/A & .35T sim & 30 & \begin{tabular}[c]{@{}l@{}}whole body:\\    \\ 29.7 ± 4.4,\\    \\ fat:\\    \\ 16.34 ± 2.67\\    \\ muscle:\\    \\ 23.36 ± 2.85\\    \\ bone:\\    \\ 105.90 ± 22.80 (HU)\end{tabular} & N/A \\ \hline
			Cusumano et al\cite{RN181} & 2020 & \begin{tabular}[c]{@{}l@{}}Abdomen\\    \\ Pelvis\end{tabular} & U-Net &  & pix2pix & .35T sim & 120 & \begin{tabular}[c]{@{}l@{}}Abdomen:\\    \\ 78.7 ± 18.5\\    \\ Pelvis:\\    \\ 54.3 ± 11.9 (HU)\end{tabular} & \begin{tabular}[c]{@{}l@{}}abdomen 2\%/2 mm:\\    \\ 98.7 ± 1.1\%\\    \\ pelvis 2\%/2 mm:\\    \\ 99.0 ± 0.7\%\end{tabular} \\ \hline
			Liu et al\cite{RN182} & 2020 & Abdomen & U-Net &  & N/A & Dixon MRI & 31 & \begin{tabular}[c]{@{}l@{}}liver: 24.1\\    \\ spleen: 28.6\\    \\ lungs: 105.7\\    \\ vertebral bodies:   110.1 (HU)\end{tabular} & \begin{tabular}[c]{@{}l@{}}mean differences   for all PTV and OAR dose metrics\\    \\ \textless .15 Gy\end{tabular} \\ \hline
			Massa et al\cite{RN148} & 2020 & Brain & Inception V3, U-Net &  & N/A & \begin{tabular}[c]{@{}l@{}}1.5T T1W, T2W, T1C,   FLAIR\\    \\ (separately)\end{tabular} & 92 & 51.2 ± 4.5 HU & N/A \\ \hline
			Andres et al\cite{RN183} & 2020 & Brain & HighResNet & \begin{tabular}[c]{@{}l@{}}Residual\\    \\ 3D\end{tabular} & N/A & 1.5T, 3.0T T1W, T1C & 402 & 92 ± 23 HU & \begin{tabular}[c]{@{}l@{}}3\%/3 mm:\\    \\ 99.8 ± 0.2\%\end{tabular} \\ \hline
			Koike et al\cite{RN149} & 2020 & Brain (GMB) & U-Net &  & pix2pix & T1W, T2W, FLAIR & 15 & \begin{tabular}[c]{@{}l@{}}whole body:\\    \\ 108.1 ± 24.0\\    \\ soft tissue:\\    \\ 38.9 ± 10.7\\    \\ bone:\\    \\ 366.2 ± 62.0 (HU)\end{tabular} & 2\%/ 2mm:   99.2 ± 1.0\% \\ \hline
			Olin et al\cite{RN184} & 2020 & H\&N & U-Net &  & N/A & Dixon MRI & 11 & \begin{tabular}[c]{@{}l@{}}Body: 94 ± 14\\    \\ Air: 300 ± 69\\    \\ Soft tissue:\\    \\ 41 ± 4\\    \\ bone:\\    \\ 258 ± 51 (HU)\end{tabular} & \begin{tabular}[c]{@{}l@{}}all dosimetric\\    \\ parameters\\    \\ within ±1\%\end{tabular} \\ \hline
			Largent et al\cite{RN142} & 2020 & H\&N & U-Net &  & GAN & T2W & 8 & 82.8 HU & N/A \\ \hline
			Qi et al\cite{RN185} & 2020 & H\&N & U-Net &  & pix2pix & \begin{tabular}[c]{@{}l@{}}T1W, T2W,\\    \\ T1C, Dixon\end{tabular} & 45 & 70.0 ± 12.0   HU & 2\%/ 2mm:   99.3 ± 0.2 \\ \hline
			Klages et al\cite{RN186} & 2020 & H\&N & CycleGAN &  & GAN & mDixon FFE & 23 & \begin{tabular}[c]{@{}l@{}}pix2pix: 66.9±7.3\\    \\ CycleGAN: 82.3±6.4   (HU)\end{tabular} & Absolute percent   mean/max dose errors   \textless 2\% \\ \hline
			Tie et al\cite{RN187} & 2020 & \begin{tabular}[c]{@{}l@{}}H\&N\\    \\ (nasopharynx)\end{tabular} & ResU-Net & Residual & cGAN & T1W, T1C, T2W & 32 & \begin{tabular}[c]{@{}l@{}}75.7 ± 14.6\\    \\ bone:\\    \\ 194.6 ± 38.9 (HU)\end{tabular} & N/A \\ \hline
			Bahrami et al\cite{RN188} & 2020 & Pelvis & \begin{tabular}[c]{@{}l@{}}SegNet\\    \\ (U-Net)\end{tabular} & Residual & N/A & 3.0T, T2W & 15 & \begin{tabular}[c]{@{}l@{}}30.0 ± 10.4\\    \\ HU\end{tabular} & N/A \\ \hline
			Florkow et al\cite{RN189} & 2020 & Pelvis & U-Net & 3D & N/A & 3T T1W & \begin{tabular}[c]{@{}l@{}}23,\\    \\ 17 dogs\end{tabular} & \begin{tabular}[c]{@{}l@{}}humans: 33\\    \\ dogs: 35 (HU)\end{tabular} & N/A \\ \hline
			Kazemirfar et al\cite{RN190} & 2019 & Brain & U-Net &  & GAN & 1.5T T1Gd & 77 & 47.2 ± 11.0 HU & \begin{tabular}[c]{@{}l@{}}2\%/ 2mm:\\    \\ 99.2 ± .8\%\end{tabular} \\ \hline
			Lei et al\cite{RN191} & 2019 & \begin{tabular}[c]{@{}l@{}}Brain\\    \\ Prostate\end{tabular} & FCN & Dense & CycleGAN & \begin{tabular}[c]{@{}l@{}}Brain: T1W\\    \\ Prostate: T2W\end{tabular} & 44 & \begin{tabular}[c]{@{}l@{}}Brain: 55.7\\    \\ Prostate: 50.8 (HU)\end{tabular} & N/A \\ \hline
			Liu et al\cite{RN192} & 2019 & Brain & VGG16 & Residual & N/A & 1.5T T1W & 40 & 75 ± 23 HU & \begin{tabular}[c]{@{}l@{}}The absolute\\    \\ percentage\\    \\ differences:\\    \\ PTV: 0.24   ± 0.46\%\\    \\ max dose: 1.39 ± 1.31\%\\    \\ V95:   0.27 ± 0.79\%\end{tabular} \\ \hline
			Liu et al\cite{RN274} & 2019 & Liver & FCN & Dense & cycleGAN & T1W & 21 & 72.9 ± 18.2 HU & \begin{tabular}[c]{@{}l@{}}2\%/ 2 mm\\    \\ 10\% LDT:\\    \\ 97.0 ± 2.9\%\end{tabular} \\ \hline
			Olberg et al\cite{RN193} & 2019 & Breast & FCN &  & GAN & .35T sim & 60 & 16.1 ± 3.5 HU & 2\%/ 2mm \textgreater 98\% \\ \hline
			Gupta et al\cite{RN194} & 2019 & H\&N & U-Net &  & N/A & 3T Dixon & 60 & \begin{tabular}[c]{@{}l@{}}81.0 ± 14.6\\    \\ air:\\    \\ 233.8 ± 28.0\\    \\ soft tissue:\\    \\ 17.6 ± 3.4\\    \\ bone:\\    \\ 193.1 ± 38.3 (HU)\end{tabular} & \begin{tabular}[c]{@{}l@{}}mean target dose   difference of\\    \\ 2.3 ± 0.1\%\end{tabular} \\ \hline
			Dinkla et al\cite{RN150} & 2019 & H\&N & U-Net &  & N/A & 3T T2W & 34 & 75 ± 9 HU & \begin{tabular}[c]{@{}l@{}}2\%/2mm:\\    \\ 95.6 ± 2.9\%\end{tabular} \\ \hline
			Wang et al\cite{RN195} & 2019 & NPC & U-Net &  & N/A & 1.5T T2W & 33 & \begin{tabular}[c]{@{}l@{}}whole body:\\    \\ 131 ± 24\\    \\ soft tissue:\\    \\ 97 ± 13\\    \\ bone:\\    \\ 357 ± 44 (HU)\end{tabular} & N/A \\ \hline
			Fu et al\cite{RN196} & 2019 & Pelvis & U-Net & 3D & N/A & 1.5T T1W & 20 & \begin{tabular}[c]{@{}l@{}}2D CNN:\\    \\ 40.5 ± 5.4\\    \\ 3D CNN:\\    \\ 37.6 ± 5.1 (HU)\end{tabular} & N/A \\ \hline
			Largent et al\cite{RN197} & 2019 & Prostate & U-Net &  & GAN & 3T T2W & 39 & \begin{tabular}[c]{@{}l@{}}U-Net:\\    \\ 34.4 ± 7.7\\    \\ GAN:\\    \\ 34.1 ± 7.5 (HU)\end{tabular} & \begin{tabular}[c]{@{}l@{}}1\% /1 mm,\\    \\ 10\% LDT:\\    \\ 99.2 ± 1.0\end{tabular} \\ \hline
			Emami et al\cite{RN198} & 2018 & Brain & ResNet & Residual & GAN & 1.0T T1Gd & 15 & \begin{tabular}[c]{@{}l@{}}whole body:\\    \\ 89.3 ± 10.3\\    \\ tissue:\\    \\ 41.9 ± 8.6\\    \\ Bone/Air:\\    \\ 240-255 (HU)\end{tabular} & N/A \\ \hline
			Arabi et al\cite{RN199} & 2018 & Pelvis & U-Net &  & N/A & 3T T2W & 39 & 32.7 ± 7.9 HU & \begin{tabular}[c]{@{}l@{}}1\%/1 mm:\\    \\ 94.6 ± 5.7\%\end{tabular} \\ \hline
			Chen et al\cite{RN200} & 2018 & Prostate & U-Net &  & N/A & 3T T2W & 51 & 30.0 ± 4.9 HU & 2\%/2 mm: 99.4\% \\ \hline
			Han\cite{RN201} & 2017 & Brain & U-Net &  & N/A & 1.5T T1W & 18 & 84.8 ± 17.3 HU & N/A \\ \hline
		\end{longtable}
	}
		\end{landscape}
	
	\begin{table}[]
		\centering
		\caption{MRI-based Synthetic CT Studies for Proton and Boron Therapy}
		\label{tab:my-table}
		\resizebox{\textwidth}{!}{%
			\begin{tabular}{llllllllll}
				\hline
				Study & Year & Site & \begin{tabular}[c]{@{}l@{}}Network\\    \\ Architecture\end{tabular} & \begin{tabular}[c]{@{}l@{}}Network\\    \\ Features\end{tabular} & GAN & Imaging Modalities & Patient Number & MAE & Dosimetry \\ \hline
				\begin{tabular}[c]{@{}l@{}}Zimmermann\\    \\ et al\cite{RN140}\end{tabular} & 2022 & Brain & Res U-Net & Residual & N/A & T1W, T2W, T1C & 47 & \begin{tabular}[c]{@{}l@{}}T1:\\    \\ body 79.8\\    \\ bone 216.3\\    \\ T2:\\    \\ body 71.1\\    \\ bone 186.1\\    \\ T1C:\\    \\ body 82.9\\    \\ bone   236.4 (HU)\end{tabular} & dose parameters within   1\% \\ \hline
				Zhao et al\cite{RN159} & 2022 & Brain & SARU & \begin{tabular}[c]{@{}l@{}}Attention Residual\\    \\ Boron- therapy\end{tabular} & N/A & T1W & 104 & Head:   67.8 ± 24.3 Skull: 144.0 ± 45.83 Brain:   14.9 ± 21.2 (HU) & 2\%/2 mm: 0.98 ±   0.01 \\ \hline
				Wang et al\cite{RN57} & 2022 & Brain & Res U-Net & Residual sRPSP & \begin{tabular}[c]{@{}l@{}}ccGAN\\    \\ (constant\\    \\ cycle)\end{tabular} & T1W, T2W, FLAIR & 195 & \begin{tabular}[c]{@{}l@{}}42 ± 13\\    \\  HU\end{tabular} & 10\%/3 mm:   55-60\% from chart \\ \hline
				Wang et al\cite{RN202} & 2021 & Brain & \begin{tabular}[c]{@{}l@{}}Attention\\    \\ U-Net\end{tabular} & Attention & cycleGAN & 1.5T, 3.0T T1W & 125 & 65.3±13.9 HU & \begin{tabular}[c]{@{}l@{}}mean absolute   differences:\\    \\ V95 1.1 ± 0.8\%\\    \\ 80\% beam axis   distal falloff\\    \\ 1.1±0.9 mm\end{tabular} \\ \hline
				Liu et al\cite{RN156} & 2021 & H\&N & Residual FCN & Residual, Dual   Energy CT & label GAN   (conditional cycleGAN) & 1.5T T1W & 57 & \begin{tabular}[c]{@{}l@{}}Low Energy CT:\\    \\ 80.0 ± 18.1   High Energy CT:  80.2 ± 16.3   (HU)\end{tabular} & N/A \\ \hline
				\begin{tabular}[c]{@{}l@{}}Maspero\\    \\ et al\cite{RN157}\end{tabular} & 2020 & Brain & U-Net & 2.5D & cGAN & 1.5T, 3.0T T1W & 60 & 61 ± 14 HU & \begin{tabular}[c]{@{}l@{}}2\%/2mm: photon 99.5   ± 0.8\%\\    \\ proton 99.2 ± 1.1\%\end{tabular} \\ \hline
				\begin{tabular}[c]{@{}l@{}}Kazemifar\\    \\ et al\cite{RN203}\end{tabular} & 2020 & Brain, proton & U-Net &  & GAN & 1.5T T1W & 77 & 47.2 ± 11.0   HU & \begin{tabular}[c]{@{}l@{}}mean absolute   difference:\\    \\ CTV \textless .5\%   (0.3 Gy)\\    \\ OAR \textless 2\% (1.2 Gy)\end{tabular} \\ \hline
				Florkow et al\cite{RN204} & 2020 & Wilms Tumor & U-Net & 3D & N/A & 1.5T T1W, T2W & 54 & 57 ± 12 HU & \begin{tabular}[c]{@{}l@{}}2\%/2 mm:  VMAT \textgreater{}99\%\\    \\ PBS (pencil beam   scanning) \textgreater{}96\%\end{tabular} \\ \hline
				Liu et al\cite{RN207} & 2019 & Liver & FCN & Dense & CycleGAN & T1W & 21 & 72.9 ± 18.2 HU & 1\%/1 mm: \textgreater{}99\% \\ \hline
				\begin{tabular}[c]{@{}l@{}}Shafai-Erfani\\    \\ et al\cite{RN205}\end{tabular} & 2019 & Brain & FCN & Dense & cycleGAN & 1.5 T1W & 50 & 54.6±6.8 HU & 2\%/ 2 mm, 10\% LDT:   98\% \\ \hline
				Neppi et al\cite{RN206} & 2019 & Brain & U-Net & 3D & N/A & 1.5T T1W & 89 & 137±32 HU & 2\%/ 2mm: 99.3\% \\ \hline
				Liu et al\cite{RN208} & 2019 & Pelvis & FCN & Dense & cycleGAN & 1.5 T2W & 17 & 51.3 ± 16.9 HU & \begin{tabular}[c]{@{}l@{}}2 mm/2\%:  97.95 ± 2.95\%\\    \\ mean Bragg peak   shift:\\    \\ 0.18 ± 0.07 cm\end{tabular} \\ \hline
			\end{tabular}%
		}
	\end{table}
	
	\noindent 
	\subsection{CT and CBCT-Based Synthetic MRI}
	
	Generating sMRI from CT leverages MRI’s high soft tissue contrast for improved segmentation accuracy and pathology detection for CT-only treatment planning. In addition, the ground truth X-ray attenuation information is maintained compared to an MRI-only workflow. Cone beam CT (CBCT) is primarily used for patient positioning before each fraction of radiation therapy. Kilovoltage (kV) and megavoltage (MV) energies are standard in CBCT with kV images providing superior contrast and MV images providing superior tissue penetration. However, noise and artifacts can often reduce CBCT image quality.\cite{RN209} Generating CBCT-based sMRI can yield higher image quality and soft-tissue contrast while also retaining CBCT’s fast acquisition speed. CT and CBCTs’ rapid acquisition time can make it preferable over MRI for patients with claustrophobia during the MR simulation or for pediatric patients who would require additional sedation. In addition, MRI is not suitable for patients with metal implants such as pacemakers. However, sMRI is significantly more challenging to generate compared to sCT. This is primarily due to the recovery of soft tissue structures visible only in MRI. For this reason, sMRI is often used to improve segmentation results in CT and CBCT. However, some studies report direct use of sMRI for segmentation. Since MRI intensity is only relative and not in definitive units like CT, MAE is much less meaningful than other metrics. Therefore, peak signal to noise ratio (PSNR) is preferentially reported.\cite{RN223}
	
	For CT-based sMRI, Dae et al implements a cycleGAN for sMRI synthesis with dense blocks in the generator. The sMRIs are input into MS-RCNN improving segmentation performance.\cite{RN210} Kalantar et al compares U-Net, U-Net++, and cycleGAN, concluding that cycleGAN preforms the best.\cite{RN211} Lei et al incorporates dual pyramid networks to extract features from both sMRI and CT and includes attention to achieve exceptional results.\cite{RN212} BPGAN synthesizes both sMRI and sCT bidirectionally with a cycleGAN. Pathological prior information, an edge retention loss, and spectral normalization improve accuracy and training stability.\cite{RN213} Both CBCT-based sMRI studies, from Emory’s Deep Biomedical Imaging Lab, significantly improve CBCT segmentation results. In their first paper, Lei et al generates sMRI with a CycleGAN, then inputs this into an attention U-Net.\cite{RN8} Fu et al makes additional improvements by generating the segmentations with inputs from both CBCT and sMRI and also including additional pelvic structures. Example contours overlaid onto CBCT and sMRI are shown in Figure 5.\cite{RN214}
	
	\begin{figure}
		\centering
		\noindent \includegraphics*[width=6.50in, height=4.20in, keepaspectratio=true]{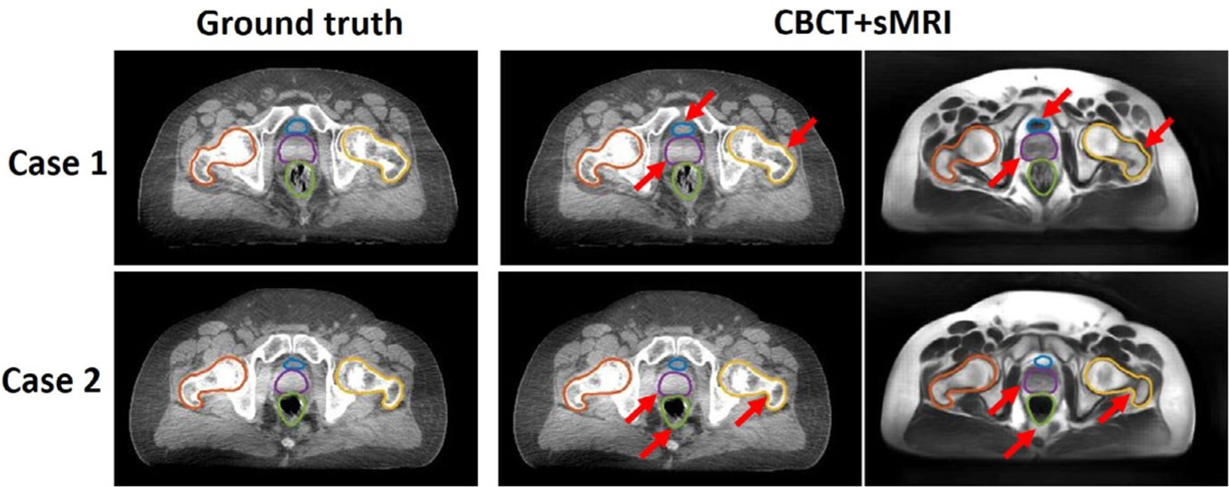}
		
		\noindent Figure 5. Contours of segmented pelvic organs for two representative patients. Ground truth contours are overlaid onto CBCT. The predicted contours of the proposed method are overlaid on CBCT and sMRI. Red arrows highlight regions in which CBCT and sMRI provide complementary information for bony structure and soft tissue segmentation. Reprinted by permission from John Wiley and Sons: Medical Physics, Pelvic multi-organ segmentation on cone-beam CT for prostate adaptive radiotherapy by Fu et al.\cite{RN214} © 2020. 
	\end{figure}
	
	\begin{table}[]
		\centering
		\caption{Synthetic MRI Studies}
		\label{tab:my-table}
		\resizebox{\textwidth}{!}{%
			\begin{tabular}{llllllllll}
				\hline
				Study & Year & Site & \begin{tabular}[c]{@{}l@{}}Network\\    \\ Architecture\end{tabular} & \begin{tabular}[c]{@{}l@{}}Network\\    \\ Features\end{tabular} & GAN & \begin{tabular}[c]{@{}l@{}}Input\\    \\  Modality\end{tabular} & \begin{tabular}[c]{@{}l@{}}Output \\    \\ Modality\end{tabular} & \begin{tabular}[c]{@{}l@{}}Patient\\    \\ Number\end{tabular} & Results \\ \hline
				Dai et al\cite{RN210} & 2021 & H\&N & MS-RCNN & \begin{tabular}[c]{@{}l@{}}Attention\\    \\ Dense\\    \\ ROI\\    \\ 3D\end{tabular} & N/A & CT & T1W & 108 & \begin{tabular}[c]{@{}l@{}}local DSC 0.77\\    \\ public DSC: 0.86\end{tabular} \\ \hline
				\begin{tabular}[c]{@{}l@{}}Kieselmann\\    \\ et al\cite{RN90}\end{tabular} & 2021 & H\&N & U-Net &  & CycleGAN & CT & 3T T2W & 27 & DSC: 0.77±0.07 \\ \hline
				\begin{tabular}[c]{@{}l@{}}Gotoh\\    \\ et al\cite{RN215}\end{tabular} & 2021 & Lumbar Spine & U-Net &  & pix2pix & CT & 3T T2W & 22 & \begin{tabular}[c]{@{}l@{}}PSNR:\\    \\ 18.4 ± 2.1\\    \\ MSE:\\    \\ 8876.7 ± 1192.9\end{tabular} \\ \hline
				\begin{tabular}[c]{@{}l@{}}Kalantar\\    \\ et al\cite{RN216}\end{tabular} & 2021 & Pelvis & U-Net &  & CycleGAN & CT & 1.5T T1W & 17 & \begin{tabular}[c]{@{}l@{}}PSNR:\\    \\ 18.3 ± 0.2\\    \\ MAE:\\    \\ 0.057 ± 0.001\end{tabular} \\ \hline
				Lei et al\cite{RN212} & 2021 & Pelvis & FCN & \begin{tabular}[c]{@{}l@{}}Attention\\    \\ Dense\end{tabular} & CycleGAN & CT & T2W & 140 & DSC: 0.95 ± 0.05 \\ \hline
				Xu et al\cite{RN213} & 2020 & Brain & FCN & \begin{tabular}[c]{@{}l@{}}Dense\\    \\ 3D\end{tabular} & N/A & CT & T1W & 391 & \begin{tabular}[c]{@{}l@{}}sMRI MAE: 15.5\\    \\  sCT MAE: 9.1\end{tabular} \\ \hline
				Li et al\cite{RN141} & 2020 & Brain & U-Net & Attention Dense & N/A & CT & 1.5T T1W & 34 & \begin{tabular}[c]{@{}l@{}}MAE: 74.2\\    \\ PSNR: 32.4\end{tabular} \\ \hline
				Fu et al\cite{RN214} & 2020 & Pelvis & U-Net & \begin{tabular}[c]{@{}l@{}}Attention\\    \\ Dense\\    \\ 3D\end{tabular} & CycleGAN & CBCT & T2W & 100 & \begin{tabular}[c]{@{}l@{}}bladder 0.96 ± 0.03\\    \\ prostate 0.91 ±   0.08 \\    \\ rectum: 0.93 ± 0.04\\    \\ (DSC)\end{tabular} \\ \hline
				Lei et al\cite{RN79} & 2020 & Pelvis & FCN & \begin{tabular}[c]{@{}l@{}}Attention\\    \\ Dense\\    \\ 3D\end{tabular} & CycleGAN & CBCT & T2W & 100 & \begin{tabular}[c]{@{}l@{}}bladder 0.95 ± 0.02\\    \\  prostate 0.86 ± 0.06\\    \\  rectum 0.91 ± 0.04\\    \\ (DSC)\end{tabular} \\ \hline
				Dong et al\cite{RN68} & 2019 & Pelvis & U-Net & 3D & CycleGAN & CT & T2W & 102 & \begin{tabular}[c]{@{}l@{}}Bladder: 0.95±0.03\\    \\  Prostate: 0.87±0.04 \\    \\ Rectum: 0.89±0.04\\    \\ (DSC)\end{tabular} \\ \hline
			\end{tabular}%
		}
	\end{table}

	\noindent 
	\subsection{Intramodal MRI Synthesis and Super Resolution}
	
	It can be beneficial to synthesize MRI sequences from other MRI sequences. Intra-modal applications include generating synthetic contrast MRI to prevent the need for injected contrast, super-resolution MRI to improve image quality and reduce acquisition time, and synthetic 7T MRI due to its lack of widespread availability and improve spatial resolution and contrast.\cite{RN217} To reduce complexity and cost, a potential approach to radiation therapy is to rotate the patient instead of using a gantry. However, the patient’s organs deform under gravity, requiring multiple MRIs at different angles for MRgRT. MR images of patients rotated at different angles can better enable gantry free radiation therapy. In this section, synthesis studies which synthesize other MRI sequences are discussed.
	
	Preetha et al synthesizes T1C images with a multi-channel T1W, T2W, and FLAIR MRI sequences using the pix2pix architecture.\cite{RN12} A cycleGAN with a ResUNet generator is trained to generate lateral and supine MR images for gantry-free radiation therapy.\cite{RN218} ResUNet is also implemented to generate ADC uncertainty maps from ADC maps for prostate cancer and mesothelioma.\cite{RN219} Studies designed explicitly for super-resolution include Chun et al and Zhao et al. In the former study, a U-Net based denoising autoencoder is trained to remove noise from clinical MRI. Since there is a limited number of paired low-resolution and high-resolution MR images, a CNN is trained to downsample high resolution data from this dataset. Finally, a GAN utilizing both residual and skip connections synthesizes the high resolution MRI with high accuracy.\cite{RN11} The same architecture is employed in Kim et al\cite{RN220} for real-time 3D MRI to increase spatial resolution. In addition, dynamic keyhole imaging is formulated to reduce acquisition time by only sampling central k-space data associated with contrast. The peripheral k-space data associated with edges is added from previously generated super-resolution images in the same position.\cite{RN220} Zhao et al makes use of super-resolution for brain tumor segmentation, increasing the dice score from 0.724 to 0.786 with 4x super resolution images generated from a GAN architecture. The generator has low- and high-resolution paths and dense blocks.\cite{RN10} Often in clinical practice, the through place resolution is increased to reduce the MRI scan time.  Xie et al achieves near perfect accuracy in recovering 1 mm from 3 mm through plane resolution by training parallel CycleGANs which predict the higher resolution coronal and sagittal slices, respectively. These predictions are then fused to create the final 3D prediction.\cite{RN284} No studies have published yet to synthesize 7T MRI for radiation therapy.
	
	\begin{table}[]
		\centering
		\caption{Intramodal MR Synthesis Studies}
		\label{tab:my-table}
		\resizebox{\textwidth}{!}{%
			\begin{tabular}{llllllllll}
				\hline
				Study & Year & Site & \begin{tabular}[c]{@{}l@{}}Network\\    \\ Architecture\end{tabular} & Network Features & GAN & Input & Output & Patients & Results \\ \hline
				Xie et al\cite{RN284} & 2022 & \begin{tabular}[c]{@{}l@{}}brain\\    \\ (BraTS)\end{tabular} & ResUnet & Residual & \begin{tabular}[c]{@{}l@{}}parallel\\    \\ CycleGANs\end{tabular} & \begin{tabular}[c]{@{}l@{}}1x1x3 mm3\\    \\ T1W, T2W, T1C,   FLAIR (separately)\end{tabular} & \begin{tabular}[c]{@{}l@{}}1x1x1 mm3\\    \\ T1W, T2W, \\    \\ T1C, FLAIR \\    \\ (separately)\end{tabular} & 300 & SSIM: 0.98 ± 0.01 \\ \hline
				Xie et al\cite{RN286} & 2022 & \begin{tabular}[c]{@{}l@{}}brain\\    \\ (BraTS)\end{tabular} & Retina U-Net & ROI & N/A & T1W & T1C & 369 & SSIM: 0.99 ± 0.01 \\ \hline
				Zhou et al\cite{RN2} & 2022 & brain & Dual path DenseNet & Dense & GAN & 3.0T T2-Flair & 4x High Res & 237 & DSC: .79 \\ \hline
				Chen et al\cite{RN218} & 2022 & pelvis & ResUnet & Residual & CycleGAN & T1W & Rotated T1W & 23 & \begin{tabular}[c]{@{}l@{}}Prone MAE:\\    \\ 35.6 ± 4.0\\    \\ Lateral MAE:\\    \\ 40.5 ± 5.8\end{tabular} \\ \hline
				Zormpas-Petridis et   al\cite{RN219} & 2022 & \begin{tabular}[c]{@{}l@{}}prostate,\\    \\ mesothelioma\end{tabular} & ResUnet &  & N/A & ADC map & ADC Uncertainty Map & 44 & ADC uncertainty   differed by 4.3\% for the prostate and 3.7\% for mesothelioma \\ \hline
				Preetha et al\cite{RN5} & 2021 & brain & U-Net &  & pix2pix & T1W, T2W, FLAIR & Synthetic Contrast & 206 & median SSIM: 0.82 \\ \hline
				Chun et al\cite{RN4} & 2019 & \begin{tabular}[c]{@{}l@{}}torso\\    \\ abdomen\end{tabular} & FCN & Residual & N/A & .35T MRI &  & 480 & SSIM: 0.96 \\ \hline
			\end{tabular}%
		}
	\end{table}

	\noindent 
	\section{RADIOMICS (CLASSIFICATION)}
	
	Unlike synthesis which maps one imaging modality to another, radiomics extracts imaging data to classify structures or to predict a value. Deep learning applications to MRI-based radiomics often achieve state-of-the-art performance over hand-crafted methods in detection and treatment outcome prediction tasks. Traditional radiomics algorithms apply various hand-crafted matrices based on shape, intensity, texture, and imaging filters to generate features. The majority of these features have no predictive power, and would confuse the model if all were directly implemented. Therefore, an important step is feature reduction which screens out features without statistical significance. Typically, this is done with a regression such as analysis of variance (ANOVA), Least Absolute Shrinkage and Selection Operator (LASSO), or ridge regression. Alternatively, a CNN or other neural network can learn significant features. The advantage of the deep learning approach is that the network can learn any relevant features including handcrafted ones. However, this assumes a large enough dataset which can be problematic for small medical datasets. Hand-crafted features have no such constraint and are easily interpretable. It is often the case that a hybrid approach including both hand-crafted and deep learning features yields the highest performance. Biometric data like tumor grade, patient age, and biomarkers can also be included as features. Once the significant features are found, supervised machine learning algorithms like support vector machines, artificial neural networks, and random forests are employed to make a prediction from these features. Recently, CNNs like Xception and InceptionResNet\cite{RN221}, recurrent neural networks with GRU and LSTM blocks, and transformers have also found favor in this task, as introduced in Section 3. Radiomics can also be done purely with deep learning as it is done with segmentation and synthesis. In this section, we divide the studies into those detecting or classifying objects in the image and studies predicting a value such as the likelihood of distant metastases, treatment response, and adverse effects. While detection is traditionally under the purview of segmentation, the architectures of detection methods and the classification task are in common with other radiomics methods, and so are discussed here.
	
	While radiomics algorithms can excel on local datasets, the main concern for MRI applications is the generalizability of the methods. Variability in MR imaging characteristics such as field strength, scanner manufacturer, pulse sequence, ROI or contour quality, and the feature extraction method can result in different features being significant. This variability can largely be mitigated by normalizing the data to a reference MRI and including data from multiple sources.\cite{RN222}
	
	Classification accuracy is an appealing evaluation metric due to its simplicity, but accuracy can be misleading with unbalanced data. For example, if 90\% of tumors in the dataset are malignant, a model can achieve 90\% accuracy by labeling every tumor as malignant. Precision\cite{RN68}, the ratio of true positives to all examples labeled as positive by the classifier, and recall\cite{RN15}, the ratio of true positives to all actual positives, will also both differ if given imbalanced data. The F1 score\cite{RN123} is defined in Eq 3, ranging from 0 to 1 and combining precision and recall to provide a single metric. A high F1 value indicates both high precision and recall and is resilient towards unbalanced data.
	
	\begin{equation} 
		F1=\frac{2(Recall*Precision)}{Recall+Precision}
	\end{equation}
	
	The most common evaluation metric resistant to unbalanced data is the area under the curve (AUC) of a receiver operating characteristics (ROC) curve\cite{RN223, RN224, RN225}. In a ROC curve, the x-axis represents the false positive (FP) rate while the y-axis relates the true positive (TP) rate. In addition, the ROC curve can be viewed as a visual representation to help find the best trade-off between sensitivity and specificity for the clinical application by comparing one minus the specificity versus the sensitivity of the model. The AUC value provides a measurement for the overall performance of the model with a value of 0.5 representing random chance and a value of 1 being perfect classification. If the AUC value is below 0.5, the classifier would simply need to invert its predictions to achieve higher accuracy. It is important to note that all these metrics are for binary classification but are commonly used in multi-class classification by comparing a particular class with an amalgamation of every other category. Finally, the concordance index (C-index) measures how well a classifier predicts a sequence of events and is most appropriate for prognostic models which predict the timing of adverse effects, tumor recurrence, or patient survival times. The C-index ranges from 0 to 1 with a value of 1 being perfect prediction.\cite{RN226, RN227} A full discussion of evaluation metrics for classification tasks is found in Hossin and Suliaman.\cite{RN228}
	
	\noindent 
	\subsection{Cancer Detection and Staging}
	
	Effectively detecting and classifying tumors is vital for treatment planning. Deep learning detection methods supersede segmentation algorithms when the tumors are difficult to accurately segment or cannot easily be distinguished from other structures. In addition, detection models can further improve segmentation results by eliminating false positives. When applied to MRI, detection studies also have the potential to differentiate between cancer types and tumor stage to potentially avoid unnecessary invasive procedures like biopsy.
	
	The majority of works in detection are for brain lesion classification. Chakrabarty et al attains exceptional results in differentiating between common types of brain tumors with a 3D CNN and outperforms traditional hand-crafted methods.\cite{RN223} Radiation induced cerebral microbleeds appear as small dark spots in 7T time of flight magnetic resonance angiography (TOF MRA) and can be difficult to distinguish from look-a-like structures. Chen et al utilizes a 3D ResNet model to differentiate between true cerebral microbleeds and mimicking structures with high accuracy.\cite{RN229} Gustafsson et al demonstrates that prostate RT DICOM structures can be accurately labeled on MRI-based sCT with InceptionResNetV2.\cite{RN123} Finally, Gao et al distinguishes between radiation necrosis and tumor recurrence for gliomas, significantly outperforming experienced neurosurgeons with a CNN.\cite{RN230}
	
	\begin{table}[]
		\centering
		\caption{Cancer Detection and Staging Studies}
		\label{tab:my-table}
		\resizebox{\textwidth}{!}{%
			\begin{tabular}{lllllllll}
				\hline
				Study & Year & Purpose & Site & Architecture & \begin{tabular}[c]{@{}l@{}}Network\\    \\ Features\end{tabular} & Input Modality & \begin{tabular}[c]{@{}l@{}}Patient\\    \\ Number\end{tabular} & Results \\ \hline
				Yang et al\cite{RN6} & 2022 & \begin{tabular}[c]{@{}l@{}}false positive\\    \\ segmentation   reduction\end{tabular} & Brain & Siamese network,   SVM & Residual & T1C & 242 & AUC: 0.93 \\ \hline
				Liang et al\cite{RN231} & 2022 & NPC Staging & Brain & FCN & \begin{tabular}[c]{@{}l@{}}Attention\\    \\ Residual\end{tabular} & T1C & 320 & AUC: 0.88 \\ \hline
				Gustafsson et al\cite{RN232} & 2022 & prostate RT DICOM   structure classification & Prostate & InceptionResNetV2 & Residual & sCT & 40 & F1: 0.985 \\ \hline
				Chakrabarty et   al\cite{RN223} & 2021 & \begin{tabular}[c]{@{}l@{}}brain tumor\\    \\ classification\end{tabular} & Brain & CNN & \begin{tabular}[c]{@{}l@{}}BRATS\\    \\ 3D\end{tabular} & T1C & 2105 & \begin{tabular}[c]{@{}l@{}}Internal:\\    \\ 0.85-100\\    \\ External: 0.73-0.99\\    \\ (AUC)\end{tabular} \\ \hline
				Gao et al\cite{RN230} & 2020 & Tumor Recurrence or   Necrosis & Brain & CNN &  & T1W, T1C, T2W & 146 & AUC: 0.96 \\ \hline
				Zhang et al\cite{RN233} & 2020 & \begin{tabular}[c]{@{}l@{}}suspected lesion\\    \\ classification\end{tabular} & Brain & Faster R-CNN,   RUSBooster &  & T1W & 121 & AUC: 0.79 \\ \hline
				Zhou et al\cite{RN234} & 2020 & \begin{tabular}[c]{@{}l@{}}brain metastases\\    \\ classification\end{tabular} & Brain & CNN &  & T1W & 266 & sensitivity: 0.81 \\ \hline
				Chen et al\cite{RN229} & 2019 & \begin{tabular}[c]{@{}l@{}}cerebrial   microbleeds\\    \\ classification\end{tabular} & Brain & ResNet & \begin{tabular}[c]{@{}l@{}}Residual\\    \\ 3D\end{tabular} & \begin{tabular}[c]{@{}l@{}}7T TOF,\\    \\ TOF-SWI MRI\end{tabular} & 73 & AUC: 0.97 \\ \hline
			\end{tabular}%
		}
	\end{table}

	\noindent 
	\subsection{Treatment Response}
	
	The decision to treat with radiation therapy is often definitive. Since radiation dose will unavoidably been delivered to healthy tissue, treatment response and the risk of adverse effects are heavily considered. Further compounding the decision, dose to healthy tissue is cumulative that is complicating any subsequent treatments. In addition, unknown distant metastasis can derail radiation therapy’s curative potential. Therefore, predicting treatment response and adverse effects are of high importance, and significant work has gone into applying deep learning algorithms to prognostic models.
	
	Diffusion weighted imaging (DWI) has attracted strong interest in studies which predict the outcome of radiation therapy. DWI measures the diffusion of water through tissue often yielding high contrast for tumors. Cancers can be differentiated by altering DWI’s sensitivity to diffusion with the b value, in which higher b values correspond to an increased sensitivity to diffusion. By sampling at multiple b-values, the attenuation of the MR signal can be measured locally in the form of apparent diffusion coefficient (ADC) values. A drawback of DWI is that the spatial resolution is often significantly worse than T1W and T2W imaging.\cite{RN235} Unlike segmentation and synthesis which require highly accurate structural information, high spatial resolution is not necessary for treatment outcome prediction, so the functional information from DWI is most easily exploited in predictive algorithms.
	
	The majority of studies seek to predict treatment outcomes and tumor recurrence. An Xception based model for predicting laryngeal and hypopharyngeal cancer local recurrence with DWI achieves good results in Tomita et al.\cite{RN236} Zhu et al takes the interesting approach of concatenating DWI histograms across twelve b values to create a “signature image.” A CNN is then applied to the signature image to achieve exceptional performance in predicting pathological complete response.\cite{RN14} Jing et al, in addition to MRI data includes clinical data like age, gender, and tumor stage to improve predictive performance.\cite{RN237} Keek et al achieves better results in predicting adverse effects by combining hand-crafted radiomics and deep learning features.\cite{RN15} Other notable papers include Huisman et al which uses an FCN suggesting that radiation therapy accelerates brain aging by 2.78 times\cite{RN238}, Hua et al which predicts distant metastases with an AUC of 0.88,\cite{RN239} and Jalalifar et al which achieves excellent results by feeding in clinical and deep learning features into an LSTM model\cite{RN240} An additional study by Jalalifar et al finds the best performance for local treatment response prediction using a hybrid CNN-transformer architecture when compared to other methods. Residual connections and algorithmic hyperparameter selection further improve results.\cite{RN241}
	
\begin{table}[]
	\centering
	\caption{Treatment Response Studies}
	\label{tab:my-table}
	\resizebox{\textwidth}{!}{%
		\begin{tabular}{lllllllll}
			\hline
			Study & Year & Purpose & Site & \begin{tabular}[c]{@{}l@{}}Network\\    \\ Architecture\end{tabular} & Network Features & \begin{tabular}[c]{@{}l@{}}Input\\    \\ Modality\end{tabular} & \begin{tabular}[c]{@{}l@{}}Patient\\    \\ Number\end{tabular} & Results \\ \hline
			Huisman et al\cite{RN238} & 2022 & Post-radiation   brain aging rate & Brain & FCN &  & 3T T1W & 32 & \begin{tabular}[c]{@{}l@{}}accelerated aging   rate:\\    \\ 2.78 years/year\end{tabular} \\ \hline
			Keek et al\cite{RN8} & 2022 & adverse reaction   prediction & Brain & xception, xgboost &  & T1Gd & 1641 & \begin{tabular}[c]{@{}l@{}}AUC: 0.71\\    \\ recall: 0.80\end{tabular} \\ \hline
			Jalalifar et al\cite{RN240} & 2022 & local tumor control   prediction & Brain & InceptionResNet +   LSTM + Clinical Feature Fusion & Residual Recurrent & T1Gd, T2-Flair & 124 & AUC: 0.86 \\ \hline
			\begin{tabular}[c]{@{}l@{}}Jalalifar\\    \\ et al\cite{RN241}\end{tabular} & 2022 & \begin{tabular}[c]{@{}l@{}}Local metastases   treatment \\    \\ response\end{tabular} & Brain & Hybrid   CNN-Transformer & \begin{tabular}[c]{@{}l@{}}3D\\    \\ Transformer\\    \\ Residual\end{tabular} & T1W, T2-FLAIR & 124 & AUC: 0.91 \\ \hline
			\begin{tabular}[c]{@{}l@{}}Hua\\    \\ et al\cite{RN239}\end{tabular} & 2022 & Distant Metastases   Prediction & H\&N & xception &  & 1.5T T1W & 441 & AUC: 0.88 \\ \hline
			\begin{tabular}[c]{@{}l@{}}Tomita\\    \\ et al\cite{RN236}\end{tabular} & 2022 & laryngeal and   hypopharyngeal cancer local recurrence prediction & H\&N & xception &  & 1.5T DWI & 70 & AUC: 0.77 \\ \hline
			\begin{tabular}[c]{@{}l@{}}Ottens\\    \\ et al\cite{RN242}\end{tabular} & 2022 & \begin{tabular}[c]{@{}l@{}}DCE-MRI   hysiological parameter estimation for tracer-kinetic\\    \\  modeling\end{tabular} & Pancreas & GRU & Recurrent & 3T DCE-MRI & 28 & \begin{tabular}[c]{@{}l@{}}random error   reduced by\\    \\  factor of 4.8\end{tabular} \\ \hline
			\begin{tabular}[c]{@{}l@{}}Zhu\\    \\ et al\cite{RN7}\end{tabular} & 2022 & rectal cancer   treatment response & Rectum & CNN &  & 3T DWI & 472 & AUC: 0.93 \\ \hline
			\begin{tabular}[c]{@{}l@{}}Zhang\\    \\ et al\cite{RN243}\end{tabular} & 2021 & Distant Metastases   Prediction & H\&N & ResNet, clinical,   regression & Residual & T2W, T1C & 189 & AUC: 0.80 \\ \hline
			\begin{tabular}[c]{@{}l@{}}Jing\\    \\ et al\cite{RN237}\end{tabular} & 2021 & NPC Risk Score   Prediction & H\&N & DenseNet + clinical   data & \begin{tabular}[c]{@{}l@{}}Dense\\    \\ 3D\end{tabular} & T1W,T2C, clinical   data & 1846 & C-index: 0.67 \\ \hline
			\begin{tabular}[c]{@{}l@{}}Jang\\    \\ et al\cite{RN244}\end{tabular} & 2021 & rectal cancer   pathological response & Rectum & ShuffleNet, LSTM & Recurrent & T2W & 466 & \begin{tabular}[c]{@{}l@{}}pCR: 0.76\\    \\ Good Response: 0.72   (AUC)\end{tabular} \\ \hline
			Jin et al\cite{RN245} & 2021 & treatment response & Rectum & CNN & 3D & T1W,T2W, T1C,DWI & 622 & \begin{tabular}[c]{@{}l@{}}cohort 1: 0.95\\    \\ cohort 2: 0.92\\    \\ (AUC)\end{tabular} \\ \hline
			\begin{tabular}[c]{@{}l@{}}Gao\\    \\ et al\cite{RN246}\end{tabular} & 2021 & sarcoma response   prediction & Whole Body & VGG-19 &  & .35T DWI & 35 & accuracy: 0.83 \\ \hline
			\begin{tabular}[c]{@{}l@{}}Metz\\    \\ et al\cite{RN247}\end{tabular} & 2020 & free water   correction for Glioblastoma Recurrence prediction & Brain & ANN &  & DTI & 35 & AUC: 0.90 \\ \hline
			\begin{tabular}[c]{@{}l@{}}Zhang\\    \\ et al\cite{RN248}\end{tabular} & 2020 & rectal cancer   treatment response prediction & Rectum & CNN &  & DKI, 3.0T T2W & 401 & \begin{tabular}[c]{@{}l@{}}pCR: 0.99\\    \\ treatment response:\\    \\ 0.70\\    \\ Tumor downstaging:\\    \\ 0.79 (AUC)\end{tabular} \\ \hline
			Fu et al\cite{RN249} & 2020 & rectal cancer   treatment response prediction & Rectum & LASSO, VGG19 &  & ADC & 43 & AUC: 0.73 \\ \hline
		\end{tabular}%
	}
\end{table}

	\noindent 
	\section{REAL-TIME AND 4D MRI}
	
	Real-time MRI during treatment has recently been made possible in the clinical setting with the creation of the MRI-LINAC. Popular models include the Viewray MRIdian (ViewRay Inc, Oakwood, OH) and the Elekta Unity (Elekta AB, Stockholm). Electron return effect (ERE), which increasing dose at boundaries with differing proton densities such as the skin at an external magnetic field, guides the architecture of these models.\cite{RN250} At higher field strengths, the ERE becomes more significant, but MR image quality increases. In addition, a higher field strength can reduce the acquisition time for real-time MRI. Therefore, a balance must be struck. Both the Elekta Unity and Viewray Mridian with 1.5T and 0.35T magnetic fields, respectively, compromise by choosing lower field strengths The Elekta Unity prioritizes image quality and real-time tracking capabilities at the expense of a more severe ERE.\cite{RN251} The MRI-LINAC has enabled an exciting new era of ART wherein anatomical changes and changes to the tumor volume can be accurately discerned and optimized between treatment fractions. In addition, unique to MRgRT, the position of the tumor can be directly monitored during treatment, potentially leading to improved tumor conformality and improved patient outcomes.\cite{RN252} 
	
	Periodic respiratory and cardiac motion are common sources of organ deformation and should be accounted for optimal dose delivery to the PTV. Tracking these motions is problematic with conventional MRI since scans regularly take approximately 2 minutes per slice leading to a total typical scan time of 20 to 60 minutes.\cite{RN253} In addition to motion restriction techniques like patient-breath hold, cine MRI accounts for motion in real-time by reducing acquisition times to 15 seconds or less. This is achieved by only sampling one (2D) or more (3D) slices with short repetition times, increasing slice thickness, and undersampling. In addition, the MR signal is sampled radially in k-space to reduce motion artifacts. Capturing a 3D volume across multiple timesteps of periodic motion is known as 4D MRI.\cite{RN254}
	
	Deep learning methods can further reduce acquisition time by reconstructing intensely undersampled cine MRI slices. In addition to reconstructing from undersampled k-space MRI sequences, several approaches further reduce acquisition time. In the first approach, cine MRI and/or k-space trajectories are used to predict the timestep of a previously taken 4D MRI. However, this method requires a lengthy 4D MRI and does not adapt to changes in the tumor volume over the course of the treatment. Additional approaches include synthesizing a larger volume than cine MRI slice captures to reduce acquisition time, predicting the deformation vector field (DVF) which relays real-time organ deformation information, or determining the 3D iso-probability surfaces of the organ to stochastically determine tumor position if real-time motion adaptation is not possible.
	
	Shown in Table 11, this category is experiencing rapid growth with majority of papers being published within the current year. Notable works include Gulamhussene et al which predicts a 3D volume from 2D cine MRI or a 4D volume from a sequence of 2D cine MR slices. A simple U-Net, introduced in Section 3, is implemented to reduce inference time. The performance degrades for synthesized slices far away from the input slices but achieves an exceptional target registration error.\cite{RN17} Nie et al instead uses autoregression and the LSTM time series modeling to predict the diaphragm position and to find the matching 4D MRI volume. Autoregression outperforms an LSTM model which could be attributed to a low number of patients.\cite{RN255} Patient motion is alternatively predicted in Terpestra et al by using undersampled 3D cine MRI to generate the DVF with a CNN with low target registration error.\cite{RN256} Similarly, Romaguera et al predicts liver deformation using a residual CNN and prior 2D cine MRI. This prediction is then input into a transformer network to predict the next slice.\cite{RN257} Driever et al simply segments the stomach with U-Net and constructs iso-probability surfaces centered about the center of mass to isolate respiratory motion. These probability distributions can then be implemented in treatment planning.\cite{RN258}
	
\begin{table}[]
	\centering
	\caption{Real Time and 4D MRI Studies}
	\label{tab:my-table}
	\resizebox{\textwidth}{!}{%
		\begin{tabular}{llllllllll}
			\hline
			Study & Year & Site & Purpose & \begin{tabular}[c]{@{}l@{}}Network\\    \\ Architecture\end{tabular} & \begin{tabular}[c]{@{}l@{}}Network\\    \\ Features\end{tabular} & Inputs & Output & \begin{tabular}[c]{@{}l@{}}Patient\\    \\ Number\end{tabular} & Results \\ \hline
			Gulamhussene et   al\cite{RN10} & 2022 & liver & 4D & U-Net &  & 2D cine MRI & 4D MRI & 20 & target registration   error: 1.2 ± 0.7mm \\ \hline
			Xiao et at\cite{RN259} & 2022 & liver & 4D & U-Net & 3D & 3D MRI, 4D MRI & High quality 4D MR & 39 & \begin{tabular}[c]{@{}l@{}}inference time:  69.3 ± 5.9 ms\\    \\ Anterior-Posterior ROI\\    \\ tracking error:\\    \\ 0.50 ± 0.55\end{tabular} \\ \hline
			Driever et al\cite{RN258} & 2022 & Stomach & Location   Probability & U-Net &  & \begin{tabular}[c]{@{}l@{}}2D T2W\\    \\ Coronal MRI\end{tabular} & \begin{tabular}[c]{@{}l@{}}3D\\    \\ iso-probability surfaces\end{tabular} & 18 & \begin{tabular}[c]{@{}l@{}}Median Standard   Deviation:\\    \\ organ deformation:\\    \\ 2.0-2.9 mm\\    \\ respiratory   deformation:\\    \\ 2.7-8.8 mm\end{tabular} \\ \hline
			Nie et al\cite{RN255} & 2022 & lung & Real Time MRI & auto-regression &  & 2D cine MRI & 4D MRI & 8 & \begin{tabular}[c]{@{}l@{}}Displacement at 8 Hz:\\    \\ autoregression 0.06 ± 0.02 mm\\    \\ LSTM 0.18 ± 0.06 mm\end{tabular} \\ \hline
			Shao et al\cite{RN9} & 2022 & heart liver & Real Time MRI & FCN &  & \begin{tabular}[c]{@{}l@{}}k-space\\    \\ trajectory, prior\cite{RN9} MRI, undersampled cine MRI\end{tabular} & 3D MRI & \begin{tabular}[c]{@{}l@{}}8 cardiac\\    \\ 9 liver\end{tabular} & \begin{tabular}[c]{@{}l@{}}13-spoke k-space   cardiac DSC:\\    \\ 0.89 ± 0.02\end{tabular} \\ \hline
			Tamura et al\cite{RN260} & 2022 & lung & Real Time MRI & CycleGAN &  & 2D cine MRI & 4DCT & 5 & 3D motion predicted 1.5 seconds in   future \\ \hline
			Wei et al\cite{RN261} & 2022 & liver & Real Time MRI & FCN &  & \begin{tabular}[c]{@{}l@{}}T1W Planning MRI,\\    \\ undersampled\\    \\ treatment MRI\end{tabular} & Treatment MRI & 3 & \begin{tabular}[c]{@{}l@{}}With 12.5\% radial   undersampling and 15\% increase in noise,\\    \\ SNR improved 4.46dB   and\\    \\ SSIM by 28\%\end{tabular} \\ \hline
				Frueh et al\cite{RN11} & 2022 & Abdomen heart & Real Time MRI & CNN &  & 2D CINE MRI & \begin{tabular}[c]{@{}l@{}}Local affinity matrices,\\    \\ segmentation\end{tabular} & 1190 (MRI) & \begin{tabular}[c]{@{}l@{}}Liver DSC: 0.95/0.96\\    \\ left ventricle DSC: 0.89/0.90\\    \\ (forward pass/backwards pass):\end{tabular} \\ \hline
				\begin{tabular}[c]{@{}l@{}}Grandinetti\\    \\ et al\cite{RN262}\end{tabular} & 2022 & liver & Real Time MRI & CNN &  & \begin{tabular}[c]{@{}l@{}}Planning Dixon MRI,\\    \\ undersampled MRI\end{tabular} & Reconstructed MRI & 3 & With 12.5\% radial   undersampling, PSNR: 34.4 \\ \hline
				Zormpas-Petridis et   al\cite{RN263} & 2021 & prostate lung & \begin{tabular}[c]{@{}l@{}}acquisition time\\    \\ reduction\end{tabular} & U-Net &  & subsampled DWI & DWI & 39 & PSNR: 55.7 \\ \hline
				Terpstra et al\cite{RN256} & 2021 & lung & Real Time MRI & FCN & 3D & 3D cine MRI & DVF & 27 & target registration   error: 1.87 ± 1.65 mm \\ \hline
				\begin{tabular}[c]{@{}l@{}}Romaguera\\    \\ et al\cite{RN257}\end{tabular} & 2020 & liver & Real Time MRI & FCN & Residual Recurrent & 3T T2W & Next slice & 85 & \begin{tabular}[c]{@{}l@{}}vessel position median accuracy:\\    \\ 0.45 mm\end{tabular} \\ \hline
				Terpstra et al\cite{RN264} & 2020 & abdomen & Real Time MRI & FCN &  & 1.5T 2D cine MRI & DVF & 135 & \begin{tabular}[c]{@{}l@{}}Standard   Reconstruction with\\    \\ undersampling   factor of 25:\\    \\ Standard method   SSIM: 0.82 ± 0.07 Deep Learning SSIM: 0.80 ± 0.08\end{tabular} \\ \hline
				Kim et al\cite{RN220} & 2019 & \begin{tabular}[c]{@{}l@{}}torso\\    \\ abdomen\end{tabular} & \begin{tabular}[c]{@{}l@{}}Real Time\\    \\ MRI\end{tabular} & FCN & Residual & .35T MRI & \begin{tabular}[c]{@{}l@{}}Higher spatial,\\    \\ temporal\\    \\ resolution\end{tabular} & 4 & SSIM: 0.89 \\ \hline
			\end{tabular}%
		}
	\end{table}
	
	\noindent 
	\section{OVERVIEW AND FUTURE DIRECTIONS}
	
	Innovations in deep learning and MRI are complementary and growing at a fast pace. Shown in Figure 2, the complexity of deep learning algorithms is rapidly increasing. New systems like the MRI-LINAC have allowed for adaptive radiation therapy and real-time MRI during treatment. Despite the successes of the studies reviewed, there are more challenges to overcome.
	
	Many challenges in deep learning applications to MRI are related to limited computational resources. While MRI offers high resolution data which complements deep learning’s big data approach, the typical 3D image size is over one gigabyte of data which means that concessions must be made to apply deep learning methods. These include downsampling the original MRI, forgoing 3D convolution, and processing the images in small patches. As field strengths increase to 7 Tesla and beyond, higher resolution images, as well as generating more powerful and expensive models, computational challenges remain ever-present despite Moore’s Law.\cite{RN265} Therefore, the task is to most efficiently utilize available computational resources. One source of innovation is the increasing optimization of hardware for computer vision. The improved hardware yields higher performance and efficiency. For example, computer vision tasks have progressed from the central processing unit (CPU) to the graphics processing unit (GPU) and often to the tensor processing unit (TPU). Neuromorphic hardware has demonstrated exceptionally high efficiency and could have applications in real-time MRgRT. Deep spiking neural networks (DSNNs), which are designed for neuromorphic hardware, can more accurately model the human brain than traditional neural networks, allowing for real-time learning and adaptation.\cite{RN266} For instance, DSNN’s adaptive capabilities could find an application in real-time MRI. Instead of only relying on local context such as in ROI methods and convolution, attention and the transformer allow for direct global context by focusing on relevant regions. Currently, hybrid CNN-transformer architectures are gaining traction by strategically placing transformer layers to improve performance while also keeping the models computationally viable with convolutional layers.\cite{RN267} In the future, it is foreseeable that pure self-attention models such as the transformer will become state-of-the-art with more powerful hardware and more efficient approaches. This trend towards attention and self-attention models is shown in Figure 2 with a growing interest in attention over the last three years. This could also partially explain the drop in studies using 3D convolution and GAN architectures since more studies are devoting their resources to attention. Another cause for fewer GANs is that nine fewer MRI synthesis studies were written in 2022 in which GANs are the current state-of-the-art method. Finally, diffusion models are an alternative to GANs, which work by gradually adding noise to an image and attempting to recreate it. Although diffusion models are computationally expensive, they can generate more realistic images than GANs and may soon find applications in super-resolution and under-sampled real-time MRI.\cite{RN268, RN269}
	
	Another challenge of deep learning applications to MRgRT is that MRgRT is still a nascent field. For example, prostate brachytherapy often uses fiducial markers designed for CT. Fiducials show exceptional contrast in CT imaging but are difficult to see on MRI and can be challenging for segmentation methods.\cite{RN111} It is likely with the maturation of the field, designed fiducial markers for MRI will see greater adoption or no longer be necessary for many applications since organ motion can be directly monitored with MRgRT.\cite{RN273} In addition, high quality public datasets often remain a roadblock. However, the growing number of yearly competitions like the BraTS challenge and public databases like The Cancer Imaging Archive (TCIA) have mitigated this effect.
	
	Deep learning has also enhanced the capabilities of MRI. sCT enables MRI to generate X-ray attenuation information, super-resolution algorithms can reduce the time of acquisition or enhance clinical MRI resolution, and synthetic contrast MRI can achieve similar results to T1C MRI with T1W and T2W sequences. These are actively being researched and should improve with time. The new frontier of MRI research is the MRI-LINAC and 7T MRI. Despite better image details at higher field strengths, the ERE increases in its severity which causes unwanted dose at air-tissue interfaces. Therefore, current MRI-LINAC models operate at 0.35 and 1.5 Tesla while diagnostic MRI is commonly at 1.5 and 3 Tesla. A solution might be to synthesize low tesla MRI to higher field strengths. Similarly, 7T MRI is gaining clinical acceptance for diagnostic imaging. Despite its greater detail, it is not readily available and is more prone to artifacts associated with high strength, non-homogenous magnetic fields making it a good candidates for synthesis algorithms.\cite{RN270} Synthesis of other modalities from MRI such as ultrasound, positron emission tomography (PET) imaging, and pathological images are also on the horizon as suitable datasets become available.
	
	A prevailing theme is the cross-pollination from different disciplines. The LSTM, GRU, and transformer models were originally developed for language translation and time series estimation but are now common in radiomics, synthesis, and segmentation. The GAN is mainly implemented in synthesis but has found applications in several segmentation architectures. Similarly, super-resolution and CT- and CBCT-based sMRI improve segmentation accuracy. It is foreseeable that MRI-based synthetic functional imaging like PET, single photon emission computed tomography (SPECT), and functional MRI (fMRI) could also improve radiomics performance. Reinforcement learning has found success in two stage segmentation networks by optimally adjusting the bounding box. Monte Carlo Dropout (MCDO) has been implemented in segmentation models to visualize uncertainty but could also be used to visualize synthesis uncertainty.\cite{RN271} Similarly, it is common to include biometric and MRI scanner manufacturer information in radiomics, and synthesis papers have recently incorporated scanner information to enhance predictions. Clinical data commonly applied to radiomics methods like patient age, prostate specific antigen (PSA) level, and biopsy data may improve MRI synthesis and segmentation methods in the future. In addition, genomics data could refine treatment plans by predicting the radiosensitivity of the patient and tumor, and enhancing the prognostic value of radiomics methods.\cite{RN272}  Another source of inspiration are the sRPSP maps applied in proton therapy which bypass sCT to give accurate attenuation information of protons. It is foreseeable to also synthesize a treatment plan or dose distribution directly from MRI without the need for sCT in an MRI-only workflow. From these developments, inter-field innovation will continue to play an important role in the development of deep learning applications to MRgRT.
	
	\noindent 
	\section{CONCLUSION}
	
	In summary, deep learning approaches to MRgRT represent the state-of-the-art in segmentation, synthesis, radiomics, and real-time MRI. These algorithms are expected to continue to improve rapidly and allow for precise, adaptive radiation therapy, and an MRI-only workflow.
	
	\noindent 
	\bigbreak
	{\bf ACKNOWLEDGEMENT}
	
	This research is supported in part by the National Cancer Institute of the National Institutes of Health under Award Numbers R01CA215718, R56EB033332, R01EB032680 and P30 CA008748.

	\noindent 
	\bigbreak
	{\bf Disclosures}
	
	The authors declare no conflicts of interest.

	\noindent 
	
	\bibliographystyle{plainnat}  
	\bibliography{arxiv}      
	
\end{document}